\documentclass [prb,aps,floatfix]{revtex4}
\usepackage{float}
\usepackage{graphicx}
\begin{document}
\baselineskip 18pt
%\large{
%--------------------------------------------------------------------
\title
{\bf Self-consistent GW combined with single site DMFT for a
Hubbard model}
\date{\today}
%--------------------------------------------------------------------
\vspace{12pt}
%--------------------------------------------------------------------
\author{K. Karlsson
\\{\it Institutionen f\"or naturvetenskap,}\\
{\it H\"ogskolan i Sk\"ovde, 54128 Sk\"ovde, Sweden}\\
%F. Aryasetiawan
%\\{\it Research Institute for Computational
%Sciences,} \\{\it AIST Tsukuba Central 2, Umezono 1-1-1, Tsukuba
%Ibaraki 305-8568 Japan}
 }
%\bigwedge
%
\begin{abstract}
We combine the single site dynamical mean field theory (DMFT)
with the non-local GW method. This is done fully self-consistently and we
 apply our formalism to a one-band Hubbard model.
Eventually at  self-consistency the {\em full} self-energy and
polarization operator of the system are retrieved. Some numerical
results, in the metallic as well as the insulator regime, are
 presented and briefly discussed.
Depending on the involved interaction (GW) parameters, substantial
changes are found when the GW self-energy is incorporated.
However,  the main point of this work
is to demonstrate the applicability of the method not
to make any strict comparison with exact results and experiments.

\end{abstract}
\maketitle
%\pacs{71.45.-d,71.45.Gm}
%
\section{Introduction}
The interest for a fundamental understanding of strongly
correlated systems has greatly increased, but still there is a
lack of a satisfactory description. On the other hand, for weakly
correlated systems the density functional theory\cite{kohn} (DFT)
within the local spin-density approximation\cite{jones} (LSDA),
however limited to ground-state properties, and the GW
approximation\cite{hedin,fa1,onida} (GWA) suitable for excited
state properties, have made a substantial contribution to the
understanding of $sp$ metals and semiconducturs. Their failure is
mainly due to a poor description of the strong on-site Coulomb
interactions among partially filled $d$ or $f$ shell electrons.
The insufficiency of the GW method has however encouraged schemes
which are all designed to treat strong on-site correlations, e.g
the LDA+U approach proposed by Anisimov and coworkers\cite{zaan}
in the early 90's, in order to treat the strong correlations
existing in the Mott insulators. There exist several similar
methods\cite{solo,zaan1,anis,kat} that are based on first
principles DFT-LSDA Hamiltonians, but the strong Coulomb
interaction for electrons residing in the localized orbitals are
explicitly taken care of via a set of Hubbard like {\it
parameters}, describing static or dynamically self-energy effects.
Obviously, there is a necessity to introduce in all the LDA+U
related methods, a so called double counting correction term for
the correlated orbitals\cite{anis,kat1}.

Recently, the dynamical mean-field theory\cite{rev2,pru} (DMFT)
has been found to be very successful in the treatment of strongly
correlated electronic systems. It is a nonperturbative method and
has been used intensively for various physical properties
\cite{fuji}, such as the famous paramagnetic (PM) Mott-Hubbard
metal-insulator transition in transition metals, superconducting
cuprates, fullerene compounds as well as organic conductors. The
DMFT method becomes exact in the limit of infinite spatial
dimensionality, and maps the original lattice problem onto an
interacting {\it dynamical} impurity problem, which must be solved
self-consistently due to its implicit coupling to the surrounding
lattice. In the single site DMFT, there is a shortage of
momentum-dependent or short-range correlations, implying a purely
{\it local} (on-site) self-energy. In the context of spatial
ordering and spectral properties that vary across the Brillouin
zone,
%, such as the quasiparticle dispersion revealed in a
%photo-emission experiment,
non-local effects would of course be crucial. Significant efforts
have been made to extend the single site DMFT, to the case where
the
self-energy\cite{maier,cluster1,cluster2,cluster3,cluster4,cluster5,
edmft1,edmft2,edmft3,carter}
exhibits finite-range interactions. The single impurity model is
replaced by a cluster-impurity\cite{maier}, giving rise to
short-range correlations ranging to the boundary of the given
cluster. The general idea is that the cluster captures, albeit the
finite correlation length, the correlations within the original
infinite lattice. Some of the approaches, however, breaks the
translationally invariant nature of the original problem, a
scenario not present in the single site DMFT. The corresponding
impurity problem is considerable harder to solve with the
increased number of local degrees of freedom. Present techniques
are based on the non-crossing approximation\cite{nca,nca1} (NCA),
the iterative perturbation theory\cite{rev2} (IPT), the Quantum
Monte Carlo method\cite{rev1} or exact
diagonalization\cite{werner,rev2}. An interpolative
approach\cite{savr1} has recently been suggested, where a simple
pole expansion of the self-energy is used and the unknown
parameters entering is determined using a chosen set of
constraints.

More recent and probably one of the most promising {\it first
principles} scheme is the so called "LDA+DMFT"
approach\cite{anis1,kat,anis2}, despite the fact that the
interaction term for the localized electrons still has to be
parametrized and the double counting term remains. The parameters
are, at least in principle, obtainable from an independent
calculation such as e.g the constrained LDA
method\cite{jepsen,hyber,martin} or from experimental data. Note
that the screening in the system is not determined from first
principles. The feasibility of the approach has indeed been
demonstrated in the pioneering work by Savrasov and coworkers in
the case Plutonium (Pu)\cite{savr2} and more recently in a number
of other cases\cite{poter,evap}.

It is generally believed that the GWA quite adequately describes
the long-range part of the screening.  Short-range correlations,
on the other hand, is not taken into account properly by the
random phase approximation\cite{pines} (RPA), however captured by
the DMFT approach.
 Contrary to DMFT, the GWA is a perturbative method. The
self-energy is given by $\Sigma = GW$, where $W$ is the screened
Coulomb interaction and $G$ is the full Green's function. The
frequently used RPA screening ($W=W_{0}$) and the zeroth order
Green's function ($G=G_{0}$) provide quasi-particle spectra of
most semiconductors and insulators as well as  bandgaps in good
agreement with experiment\cite{fa1}. However, there is the
important issue of
self-consistency\cite{sc1,sc2,sc3,sc4,sc5,sc6,sc7,sc8,sc9}. If the
GWA should be conserving\cite{baym}, the self-energy requires the
Green's function as well as the the screened interaction to be
evaluated self-consistently.

The aim of this paper is to combine, fully self-consistently, the
GW method with the single site DMFT, and present numerical results
for a one-band Hubbard model. The "DMFT+GW" approach, recently
proposed by Biermann {\it et al}\cite{bier1}, includes no
Hubbard-like (parameter) interaction and consequently there is no
need for the ambiguous double counting term. The main idea is that
the large on-site part of the self-energy is calculated using DMFT
and the off-site (long-range) contribution is taken from the GWA.
We will present results using a various degrees of
self-consistency for the $GW$ self-energy. Related work along this
line can be found in Refs. \cite{gw1,gw2}.

We will study two sites per unit cell in one (1D chain) and two
dimensions (2D plane), in order to be able to study the formation
and stability of different magnetic structure.  We solve the
single site impurity problem using the exact diagonalization
method\cite{werner}. In addition to the impurity self-energy, a
two-particle correlation\cite{bier1} function is calculated,
needed for the evaluation of the {\it impurity} screened
interaction. Thus, the iterative loop will include two quantities
to be determined self-consistently: the bath Green's function
${\cal G}$ as well as the bath effective interaction ${\cal U}$.
We like to stress that the effective Hubbard ${\cal U}$ is not a
parameter, it is in fact found self-consistently.

In Sec. II. we describe the method for the calculations. In Sec.
III we present and discuss the results, and in Sec. IV we give a
short summary.

\section{Theory}
\subsection{Single site DMFT}
In this section we establish the necessary concepts and formalisms
for the so called single site DMFT, a scheme that later on is
combined with the GWA. We will consider the Hubbard model with an
on-site interaction $U$ and nearest neighbor hopping $t$ ($t=1$
while leaving the on-site interaction $U$ variable). The unit cell
will contain two sites, but a generalization is
straightforward\cite{fleck}.
 The model and the corresponding Green's function reads
\begin{eqnarray}\label{hubb}
\hat H &=&
-t\sum_{mi,nj,\sigma}a^{\dagger}_{mi\sigma}a_{nj\sigma}+
U\sum_{mi}n_{mi\uparrow}n_{mi\downarrow}
\end{eqnarray}
\begin{eqnarray}
G_{mi,nj,\sigma}(\tau)= -\theta(\tau)\langle a_{mi\sigma}(\tau)
a^{\dagger}_{nj\sigma}(0)\rangle+\theta(-\tau)\langle
a^{\dagger}_{nj\sigma}(0)a_{mi\sigma}(\tau)\rangle.
\end{eqnarray}
We define positions in the lattice by ${\bf R}_{mi}= {\bf
T}_{m}+{\bf \tau}_i$, where ${\bf \tau}_i$ labels sites within the
unit cell and $m$ a particular unit cell \cite{note1}. Using the
equation of motion for $G$ (with operator $\hat K=\hat H-\mu \hat
N$) and assuming a {\em local} self-energy,
$\Sigma_{mi,nj,\sigma}=\Sigma_{i\sigma}\delta_{ij}\delta_{mn}$,
one can show that
\begin{eqnarray}\label{Gminus}
G^{-1}_{ij\sigma}({\bf k};i\nu_n) &=& (i\nu_n + \mu)\delta_{ij} +
h_{ij}({\bf k}) - \Sigma_{i\sigma}(i\nu_n)\delta_{ij}
\end{eqnarray}
where the kinetic energy matrix is given by\cite{note2}
\begin{equation}\label{hopp}
h({\bf k}) = \pmatrix{ 0 &  2(cosk_x+cosk_y) \cr
                     2(cosk_x+cosk_y) &  0 }. \;\;\;
\end{equation}
We have also defined the real space transforms as
\begin{eqnarray}
G_{ij\sigma}({\bf k};i\omega_l)=\frac{1}{N}\sum_{n}\sum_{m}
e^{-i{\bf k}\cdot{\bf T}_{mi}}
G_{mi,nj,\sigma}(i\omega_l)e^{i{\bf k}\cdot{\bf T}_{nj}}
\end{eqnarray}
\begin{eqnarray}
G_{mi,nj,\sigma}(i\omega_l)=\frac{1}{N}\sum_{{\bf k}} e^{i{\bf
k}\cdot{\bf T}_{mi}}G_{ij\sigma}({\bf k};i\omega_l) e^{-i{\bf
k}\cdot{\bf T}_{nj}}
\end{eqnarray}
where the lattice has $N$ unit cells. In the Matsubara
formulation, we adopt the definition
\begin{eqnarray}
{\cal G}(i\nu_n) &=& \int_{0}^\beta d\tau\; e^{i\nu_n \tau}
\cal{G}(\tau)
\nonumber\\\\
\cal{G}(\tau) &=& \frac 1 \beta \sum_n e^{-i\nu_n \tau} {\cal
G}(i\nu_n)
\end{eqnarray}
where $\nu_n$ denotes the Matsubara (odd) frequency for fermion
propagators. For bosons we use $\omega_n$ (even) as a convention.
\begin{equation}\label{fermion}
\nu_n = \frac{(2n+1)\pi}{\beta},
\end{equation}
\begin{equation}\label{boson}
\omega_n=\frac{2n\pi}{\beta}.
\end{equation}

Inversion of Eq. (\ref{Gminus}) gives the lattice Green's function
%\begin{equation}
%G^{-1}({\bf k},i\nu_n) = \pmatrix{ i\nu_n + \mu -
%\Sigma_{1\sigma}(i\nu_n)& 2(cosk_x+cosk_y) \cr 2(cosk_x+cosk_y) &
%i\nu_n + \mu - \Sigma_{2\sigma}(i\nu_n)  }. \;\;\;
%\end{equation}
%has the inverse
\begin{equation}\label{Gmat}
G({\bf k},i\nu_n) = \frac{1}{D({\bf k},i\nu_n)}\pmatrix{ i\nu_n +
\mu - \Sigma_{2\sigma}(i\nu_n)& -2(cosk_x+cosk_y) \cr
-2(cosk_x+cosk_y) & i\nu_n + \mu - \Sigma_{1\sigma}(i\nu_n)  }
\;\;\;
\end{equation}
where $D({\bf k},i\nu_n)=(i\nu_n + \mu -
\Sigma_{2\sigma}(i\nu_n))(i\nu_n + \mu -
\Sigma_{1\sigma}(i\nu_n))-4(cosk_x+cosk_y)^{2}$. The local
(impurity) Green's function is calculated using the diagonal
elements;
\begin{eqnarray}\label{locsig}
G_{i\sigma}(i\nu_n)= \frac{1}{N}\sum_{{\bf k}}G_{ii\sigma}({\bf
k};i\nu_{n}).
\end{eqnarray}
Regarding the corresponding self-energy, we remark that in the
case of single site DMFT¨, no causality problems occurs, the
lattice self-energy is identical to the impurity self-energy:
$\Sigma_{ij\sigma}({\bf
k};i\nu_n)=\Sigma_{i\sigma}(i\nu_n)\delta_{ij}$.
 At DMFT self-consistency the Green's
function calculated using Eqs. (\ref{Gmat}-\ref{locsig}) must
coincide with the one extracted from the impurity model.
 We have determined the site and spin dependent
impurity Green's function using the exact
diagonalization\cite{werner} (ED) Lanczos method for the single
impurity Anderson model. In the present case (zero temperature;
$\beta \rightarrow \infty$), we have solved an effective impurity
model for each site $i=1,2$, given by
\begin{eqnarray}\label{and1}
  H_i &=& \sum_{\sigma} [ \varepsilon_d n_{i\sigma}+
  \sum^{N_{s}-1}_{k=1}   \varepsilon_{ik\sigma}
  c^{\dagger}_{k\sigma}c_{k\sigma}  +
 \sum^{N_{s}-1}_{k=1}
   V_{ik\sigma} (c^{\dagger}_{k\sigma} d_{i\sigma} + d^{\dagger}_{i\sigma}
   c_{k\sigma})]
\nonumber \\
 &+&Un_{i\uparrow}n_{i\downarrow}
\end{eqnarray}
where $ \varepsilon_d =-\mu$ is the energy  of the localized level
on the impurity site. The second term gives the energy of all the
bath (conduction band) electrons, which are labelled by
$k=1,..,N_{s}-1$. The hopping between the bath states and the
impurity state is described by the third term, where
$V_{ik\sigma}$ is a hopping matrix element.

The DMFT approach maps the original lattice problem defined by the
Hubbard model onto a self-consistent solution of the Dyson
equation in Eq. (\ref{Gmat}) and the (auxiliary) impurity problem
defined by the bath Green's function
\begin{eqnarray}\label{scg}
{\cal G}^{-1}_{i\sigma}(i\nu_n)= G^{-1}_{i\sigma}(i\nu_n)
+\Sigma_{i\sigma}(i\nu_n).
\end{eqnarray}

In order to initialize the iterations it is sufficient to guess
the parameters of the Anderson model, $\varepsilon_{ik\sigma}$ and
$V_{ik\sigma}$, as well as the bath Green's function. We construct
\begin{eqnarray}
{\cal G}_{i\sigma}(i\nu_n) &=& \frac{1}{N}\sum_{{\bf k}}[(i\nu_n +
\mu)\delta_{ij} + h_{ij}({\bf k}) -
B_{i\sigma}(i\nu_n)\delta_{ij}]^{-1}
\end{eqnarray}
where $B_{i\sigma}$ is a chosen suitable external field (in the
PM case $B_{i\sigma}=0$). Solving the effective impurity
model we derive the self-energy $\Sigma_{i\sigma} = {\cal
G}^{-1}_{i\sigma}- G^{-1}_{i\sigma}$ and proceed with the
inversion of the matrix in Eq. (\ref{Gminus}). Finally  we update
the bath Green's function using ${\cal G}_{i\sigma}(i\nu_n)=
[1/\sum_{{\bf k}} G_{ii\sigma}({\bf k};i\nu_{n})/N
+\Sigma_{i\sigma}(i\nu_n)]^{-1}$ and mix it with the previous one.
The bath ${\cal G}^{-1}$ is represented by the $U=0$ impurity
Green's function:
\begin{eqnarray}
{\cal G}^{-1}_{i\sigma}(N_s,i\nu_n) &=& (i\nu_n + \mu)  -
\sum^{N_s - 1}_{k=1}
\frac{V^{2}_{ik\sigma}}{i\nu_n-\varepsilon_{ik\sigma}}
\end{eqnarray}
 in order to provide us with a new set Anderson parameters, found
by a fitting procedure. The best choice is found by minimizing the
function\cite{werner,capone}
\begin{eqnarray}
\chi^{2}_{i\sigma}  = \frac{1}{N_{w}+1}\sum^{N_w}_{n=0} |{\cal
G}^{-1}_{i\sigma}(N_s,i\nu_n) - {\cal
G}^{-1}_{i\sigma}(i\nu_n)|/\nu_n
\end{eqnarray}
for each site $i$ and spin channel $\sigma$. The convergency with
respect to $N_s$ is very fast. We found that already $N_{s}-1 = 7$
bath states is sufficient  to describe the continuum of conduction
states. The DMFT cycle is now closed: at hand we have a new set of
Anderson parameters (which defines the impurity problem) as well
as an updated bath ${\cal G}$. At self-consistency, the Green's
function from the impurity problem should be equal to one obtained
from summing the momentum-dependent lattice Green's function over
the Brillouin zone, as done in Eq. (\ref{locsig}).

When DMFT is combined with the GWA, the the impurity charge
response is entering the formalism. The two-particle response is
defined by
\begin{eqnarray}\label{chargeres}
\chi_{i} (\tau) &=& - \langle T_{\tau} [\hat \rho_{i}(\tau) \hat
\rho_{i} ] \rangle \nonumber\\
 &=& -\langle \hat \rho_{i}(\tau) \hat
\rho_{i} \rangle \theta(\tau) - \langle \hat \rho_{i} \hat
\rho_{i}(\tau) \rangle \theta(-\tau)
\end{eqnarray}
 where $\hat \rho_{i}(\tau) \equiv \hat n_{i}(\tau)
-n_i$, $\hat \rho_{i}(\tau) = e^{\hat H_i \tau}\hat
\rho_{i}e^{-\hat H_i \tau}$ and the total charge on the impurity
is denoted by $n_i=n_{i\uparrow}+n_{i\downarrow}$. From a
numerically point of view, the charge response is evaluated on the
same footing as the Green's function, with the aid of Lanzcos
algorithm. All calculations are done for a fix chemical potential
$\mu$. The total number of electrons in the cell
\begin{eqnarray}
n= \frac{1}{2}\sum_{i\sigma} n_{i\sigma}
\end{eqnarray}
 is then allow to adjust self-consistently. In the PM case (no
 doping) $n_{i\sigma}=1/2$ for all sites and spin-channels.

\subsection{DMFT combined with the GWA}

We now consider a scheme\cite{bier1} that properly adds the
momentum-dependent GW self-energy to the local DMFT self-energy,
giving rise to a lattice self-energy which describes, in addition
to local effects, also long-range correlations. The RPA will be
used for the screened interaction, implying $\Sigma^{GW} =
G(1-UP^{GW})^{-1}U$, where we used  $v({\bf r}-{\bf
r}')=U\delta({\bf r}-{\bf r}')$
 for the bare Coulomb interaction. Note that even if $v$ is short-ranged,
 $W$ can have off-site components coming from $P^{GW}$.

 The polarization operator (bubble) in the GWA is given by
\begin{eqnarray}
P^{GW}_{ij}({\bf q};i\omega_m) &=&
\sum_{\sigma}P^{GW}_{ij\sigma}({\bf q};i\omega_m)
\end{eqnarray}
where
\begin{eqnarray}\label{Pij}
P^{GW}_{ij\sigma}({\bf q};i\omega_m) &=&
\frac{1}{\beta}\sum_{n}\frac{1}{N_k}\sum_{{\bf
k}}G_{ij\sigma}({\bf q}+{\bf
k};i\omega_{m}+i\nu_{n})G_{ji\sigma}({\bf k};i\nu_{n}).
\end{eqnarray}
The sum over ${\bf k}$ includes $N=N_k$ points in the first
Brillouin zone (BZ), and ${\bf q}$ belongs to the irreducible BZ.
The Green's function in Eq. (\ref{Pij}) is obtained by inverting
the matrix
\begin{eqnarray}
G^{-1}_{ij\sigma}({\bf k};i\nu_n) &=& (i\nu_n + \mu)\delta_{ij} +
h_{ij}({\bf k}) - \Sigma_{ij\sigma}({\bf k};i\nu_n)
\end{eqnarray}
where $\Sigma_{ij\sigma}({\bf k};i\nu_n)$ is the proper lattice
self-energy (see Eq. (\ref{totsig})). In the first iteration,
however, the local impurity self-energy $\Sigma_{i\sigma}(i\nu_n)$
is used.

 The screened interaction fulfills
$W=v+vXv=\epsilon^{-1}v$, which in the case $v({\bf r}-{\bf
r}')=U\delta({\bf r}-{\bf r}')$, using $\epsilon=1-vP^{GW}$
transforms to
\begin{eqnarray}\label{Wij}
W_{ij}({\bf q};i\omega_m) &=& U\Pi_{ij}({\bf q};i\omega_m)
\end{eqnarray}
where
%\begin{eqnarray}
%\chi_{ij}({\bf q};i\omega_m) &=& \sum_{l}A_{il}P^{GW}_{lj}({\bf
%q};i\omega_m)
%\end{eqnarray}
$\Pi$ is the matrix obtained by inverting the dielectric matrix
$[\delta_{ij}-UP^{GW}_{ij}({\bf q};i\omega_n)]$. If the Coulomb
interaction $v$ takes into account nearest ($V$) and next nearest
neighbors interaction ($V'$) the dielectric function and screened
interaction reads in 1D and 2D case respectively
\begin{eqnarray}\label{hopp}
\epsilon &=& \pmatrix{ 1-UP^{GW}_{11}-2V'cos(2q)P^{GW}_{11} &
-UP^{GW}_{12}-2Vcos(q)P^{GW}_{22}\cr
                     -UP^{GW}_{21}-2Vcos(q)P^{GW}_{11} &
                     1-UP^{GW}_{22}-2V'cos(2q)P^{GW}_{22} }
\nonumber\\
 W &=& \pmatrix{ U\epsilon^{-1}_{11}+2V'cos(2q)\epsilon^{-1}_{11} &
U\epsilon^{-1}_{12}+2Vcos(q)\epsilon^{-1}_{11}\cr
                     U\epsilon^{-1}_{21}+2Vcos(q)\epsilon^{-1}_{22} &
                     U\epsilon^{-1}_{22}+2V'cos(2q)\epsilon^{-1}_{22} }
\nonumber\\\nonumber\\
\epsilon &=& \pmatrix{
1-UP^{GW}_{11}-4V'cos(q_x)cos(q_y)P^{GW}_{11} &
-UP^{GW}_{12}-2V(cos(q_x)+cos(q_y))P^{GW}_{22}\cr
                      -UP^{GW}_{21}-2V(cos(q_x)+cos(q_y))P^{GW}_{11}&
                     1-UP^{GW}_{22}-4V'cos(q_x)cos(q_y)P^{GW}_{22} }
\nonumber\\
 W &=& \pmatrix{ U\epsilon^{-1}_{11}+4V'cos(q_x)cos(q_y)\epsilon^{-1}_{11} &
U\epsilon^{-1}_{12}+2V(cos(q_x)+cos(q_y))\epsilon^{-1}_{11}\cr
                     U\epsilon^{-1}_{21}+2V(cos(q_x)+cos(q_y))\epsilon^{-1}_{22} &
                     U\epsilon^{-1}_{22}+4V'cos(q_x)cos(q_y)\epsilon^{-1}_{22} }
                     . \;\;\;
\end{eqnarray}
Like the polarization bubble, the screened interaction is a real
valued function on the imaginary axis (even Matsubara frequencies)
and the diagonal part ($i=j$) is positive and approaches the bare
$U$ for large $\omega_m$, implying that the correlated part
(frequency dependent) of $W$ goes to zero ($W^{c}_{ij}({\bf
q};i\omega_m) \sim \delta_{ij}/ (i\omega_m)^{2}$ when $\omega_m
\rightarrow \infty$ and $V=V'=0$).

Finally we achieve for the $GW$ self-energy\cite{Sx}
($\Sigma^{GW}_{ij\sigma}({\bf
q};i\nu_n)=Un_{i-\sigma}\delta_{ij}+\Sigma^{c}_{ij\sigma}({\bf
q};i\nu_n)$ and $W^{c}_{ij}=W_{ij}-U\delta_{ij} $)
%($\Sigma^{GW}_{ij\sigma}({\bf
%q};i\nu_n)=Un_{i-\sigma}\delta_{ij}+\Sigma^{x}_{ij\sigma}({\bf
%q})+\Sigma^{c}_{ij\sigma}({\bf q};i\nu_n)$) where (in 2D)
\begin{eqnarray}\label{siggw}
\Sigma^{c}_{ij\sigma}({\bf q};i\nu_n) &=&
-\frac{1}{\beta}\sum_{m}\frac{1}{N_k}\sum_{{\bf
k}}G_{ij\sigma}({\bf q}-{\bf
k};i\nu_{n}-i\omega_{m})W^{c}_{ij}({\bf
k};i\omega_{m})\nonumber\\
&=& -\frac{1}{\beta}\sum_{m}\frac{1}{N_k}\sum_{{\cal R}}\sum_{{\bf
k}\in {\rm IBZ}}G_{ij\sigma}({\bf q}-{\cal R}{\bf
k};i\nu_{n}-i\omega_{m})W^{c}_{ij}({\bf k};i\omega_{m})
\end{eqnarray}
where $W({\cal R}{\bf k})=W({\bf k})$ has been used. ${\cal R}$
denotes a rotation matrix corresponding to a point-symmetry
operation\cite{note3}. The particle number used for the
Hartree-Fock part
($\Sigma^{HF}_{i\sigma}=Un_{i-\sigma}\delta_{ij}$) is calculated
using the {\em impurity} Green's function, however at
self-consistency the impurity Green's function and the local one
should be identical (the ${\bf k}$ dependent lattice Green's
function summed over ${\bf k}$). The total lattice self-energy,
corrected for double counting, and to be used in the construction
of the next $G^{-1}$ can thus be written as
\begin{eqnarray}\label{totsig}
\Sigma_{ij\sigma}({\bf q};i\nu_n) &=&\Sigma^{GW}_{ij\sigma}({\bf
q};i\nu_n) - \delta_{ij}\frac{1}{N_k}\sum_{{\bf
k}}\Sigma^{GW}_{ij\sigma}({\bf k};i\nu_n)+\Sigma_{i\sigma}(i\nu_n)
\nonumber\\
&=&\Sigma^{GW}_{ij\sigma}({\bf q};i\nu_n) - \delta_{ij}\sum_{{\bf k}\in {\rm
IBZ}}\Sigma^{GW}_{ij\sigma}({\bf k};i\nu_n)w_{{\bf
k}}+\Sigma_{i\sigma}(i\nu_n)
\end{eqnarray}
where $w_{{\bf k}}$ is the weight of ${\bf k}$ in the IBZ. Note
that the local part of $\Sigma$ ($\Sigma_{i\sigma}$) is usually
much larger in magnitude than the non-local part given by
$[\Sigma^{GW} - 1/N_k\sum_{{\bf k}}\Sigma^{GW}]$.

Finally the local $G$ to be used to find the bath ${\cal G}$ via
the self-consistency relation:
\begin{eqnarray}\label{scg}
{\cal G}^{-1}_{i\sigma}(i\nu_n)= G^{-1}_{i\sigma}(i\nu_n)
+\Sigma_{i\sigma}(i\nu_n)
\end{eqnarray}
 can be written as
\begin{eqnarray}
G_{i\sigma}(i\nu_n)= \frac{1}{N_k}\sum_{{\bf k}}G_{ii\sigma}({\bf
k};i\nu_{n})=\sum_{{\bf k}\in {\rm IBZ}}G_{ii\sigma}({\bf
k};i\nu_n)w_{{\bf k}}
\end{eqnarray}
where the diagonal-elements $G_{ii\sigma}({\bf k};i\nu_{n})$ are
found from inverting
\begin{eqnarray}\label{totg-1}
G^{-1}_{ij\sigma}({\bf k};i\nu_n) &=& (i\nu_n + \mu)\delta_{ij} +
h_{ij}({\bf k}) - \Sigma_{ij\sigma}({\bf k};i\nu_n)
\end{eqnarray}
with the self-energy from Eq. (\ref{totsig}).

 In an ordinary single site DMFT calculation the impurity problem is
solved for {\em fixed} on site $U$ and only the bath ${\cal G}$ is
updated and determined self-consistency via Eq. (\ref{scg}). It is
however desirable to solve the impurity problem with an updated or
an effective Hubbard interaction. The {\em static} impurity charge
response, $\chi_{i}(i\omega_m = 0)$, is used to construct the {\em
static impurity} screened interaction and polarization
\begin{eqnarray}\label{Wimpur}
W_{i}(i\omega_m= 0) &=& {\cal U}_i + {\cal
U}_i\chi_{i}(i\omega_m= 0){\cal U}_i
\end{eqnarray}
\begin{eqnarray}
P_{i}(i\omega_m= 0) &=& {\cal U}^{-1}_i - W^{-1}_{i}(i\omega_m= 0)
\end{eqnarray}
where ${\cal U}_i$ is the effective Hubbard onsite-interaction
used for the solution of the impurity problem at site $i$. Then
the full polarization kernel can be written, using Eq.
(\ref{Pij}),
\begin{eqnarray}\label{fullP}
P_{ij}({\bf q};i\omega_m) &=& P^{GW}_{ij}({\bf
q};i\omega_m)-\delta_{ij}\frac{1}{N_k}\sum_{{\bf k}}P^{GW}_{ij}({\bf
k};i\omega_m) + P_{i}(i\omega_m= 0)
\nonumber\\
&=& P^{GW}_{ij}({\bf q};i\omega_m)-\delta_{ij}\sum_{{\bf k}\in {\rm IBZ}}
P^{GW}_{ij}({\bf k};i\omega_m)w_{{\bf k}} + P_{i}(i\omega_m = 0)
\end{eqnarray}
a relation analogous to Eq. (\ref{totsig}). Then the local
screened interaction reads
\begin{eqnarray}
 \frac{1}{N_k}\sum_{{\bf k}}W_{ii}({\bf k};i\omega_m=
0)=\sum_{{\bf k}\in {\rm IBZ}}W_{ii}({\bf k};i\omega_m= 0)w_{{\bf
k}}
\end{eqnarray}
where the diagonal-elements $W_{ii}({\bf k};i\omega_m= 0)$ are
found from inverting
\begin{eqnarray}
W^{-1}_{ij}({\bf k};i\omega_m= 0) &=&  U^{-1}_i\delta_{ij} -
P_{ij}({\bf k};i\omega_m= 0).
\end{eqnarray}
Note that here the bare $U$ is used (same for all sites $U_i =
U$). In the case when hopping to neighbors is allowed we have to
substitute the diagonal term $U^{-1}_i\delta_{ij}$ with the
inverse of the bare Coulomb matrix
\begin{equation}\label{bareV}
 \pmatrix{ U + 4V'cosk_xcosk_y &  2V(cosk_x+cosk_y) \cr
                     2V(cosk_x+cosk_y) &  U + 4V'cosk_xcosk_y }. \;\;\;
\end{equation}

Finally the updated effective interaction is found from the
self-consistency relation
\begin{eqnarray}
{\cal U}^{-1}_{i}= 1/ \sum_{{\bf k}}W_{ii}({\bf k};i\omega_m=
0)/N_k +P_{i}(i\omega_{m}=0)
\end{eqnarray}
a relation analogous to Eq. (\ref{scg}), however, only the static
value of ${\cal U}^{-1}$ is used in the solution of the impurity
problem.

The spectral function is given by
\begin{eqnarray}\label{akw}
A({\bf k};\omega)= -\frac{1}{N_{\tau}\pi}{\rm
Im}\sum_{\sigma}\sum_{ij} G_{ij\sigma}({\bf k};\omega)
\end{eqnarray}
where $N_{\tau}=2$ in the antiferromagnetic
(AF) case. In order to obtain the Green's
function on the real energy axis, we use the Pade approximation
for the self-energy $\Sigma_{ij\sigma}({\bf k};i\nu_n)
\rightarrow \Sigma_{ij\sigma}({\bf k};\omega)$.

In order to study the stability of different phases, the total
energy (per site) is calculated using
\begin{eqnarray}
E = {\rm Tr}\{h({\bf k})G({\bf k};i\nu_n)\} + \frac{1}{2}{\rm
Tr}\{\Sigma({\bf k};i\nu_n) G({\bf k};i\nu_n)\}
\end{eqnarray}
where
\begin{eqnarray}
{\rm Tr} \equiv \frac{1}{\beta}\sum_{n}\frac{1}{N_k}\sum_{{\bf
k}}\frac{1}{N_{\tau}}\sum_{i}\sum_{\sigma}
\end{eqnarray}
The hopping and self-energy matrices are given in Eqs.
(\ref{hopp},\ref{totsig}) and the the Green's function matrix is
found by inverting Eq. (\ref{totg-1}).

\subsection{Computational details}
Some care has to be taken when performing the Matsubara sums for
the polarization bubble and the self-energy in Eqs.
(\ref{Pij},\ref{siggw}). The bubble can be written as
\begin{eqnarray}\label{Pij0}
P^{GW}_{ij\sigma}({\bf q};i\omega_m) &=&
\frac{1}{\beta}\sum_{n\geq 0}\frac{1}{N_k}\sum_{{\bf k}}
[G_{ij\sigma}({\bf q}+{\bf
k};i\omega_{m}+i\nu_{n})G_{ji\sigma}({\bf k};i\nu_{n})+
\nonumber\\&& \hspace{61pt}G_{ij\sigma}({\bf q}+{\bf
k};i\omega_{m}+i\nu_{-n-1})G^{*}_{ji\sigma}({\bf k};i\nu_{n})]
\end{eqnarray}
where we have used that
$G(i\nu_{-n})=G^{*}(i\nu_{n-1})$\cite{note4}. The polarization is
real valued on the imaginary axis (even Matsubara frequencies) and
the diagonal part ($i=j$) is negative; $P^{GW}_{ij}({\bf
q};i\omega_m) \sim \delta_{ij}/ (i\omega_m)^{2}$ for large
$\omega_m$. We note that for large $n$ the first term behaves as
\begin{eqnarray}
\delta_{ij}\frac{1}{\beta}\sum_{n}\frac{1}{i(\omega_m +
\nu_n)i\nu_n}.
\end{eqnarray}
This is however not the case for the second term, but yet we find
that the following procedure is appropriate: If the Matsubara sum
is done for {\em all} frequencies on the imaginary axis the result
is $-\beta/4$ for $m=0$, otherwise zero. We have subtracted the
term $\delta_{ij}/\beta\sum_{n}1/i(\omega_m + \nu_n)i\nu_n$ in Eq.
(\ref{Pij0}) (where, of course, the sum is done for finite $n$)
and consequently added $-\delta_{ij}\beta/4$.
 The second term, however is
large whenever $(m-n-1)$ around zero, due to $G(m-n-1)$, even if
$G^{*}(n)$ is decaying for large $n$. Therefore the upper limit
for the $n$-sum in Eq. (\ref{Pij0}) is chosen to depend on $m$.
Thus we evaluate
\begin{eqnarray}
P^{GW}_{ij\sigma}({\bf q};i\omega_m)  &=&
\frac{1}{\beta}\sum^{N_{p}(m)}_{n\geq 0}\frac{1}{N_k}\sum_{{\bf
k}} [G_{ij\sigma}({\bf q}+{\bf
k};i\omega_{m}+i\nu_{n})G_{ji\sigma}({\bf k};i\nu_{n})+
\nonumber\\&& \hspace{61pt}G_{ij\sigma}({\bf q}+{\bf
k};i\omega_{m}+i\nu_{-n-1})G^{*}_{ji\sigma}({\bf k};i\nu_{n})
\nonumber\\
&-& \delta_{ij}\{\frac{1}{i(\omega_m +
\nu_n)i\nu_n}+\frac{1}{i(\omega_m + \nu_{-n-1})(i\nu_n)^{*}}\}] -
\delta(\omega_m)\delta_{ij}\beta/4.
\end{eqnarray}
where $N_{p}(m)=N_{P}+m$. The polarization is calculated as
described above for $m=0,N_{h}$, whereas for $m=N_{h}+1,N_g$ we
fit to
\begin{eqnarray}
P^{GW}_{ij}({\bf q};i\omega_m)  &=&
\delta_{ij}\frac{P^{0}}{(i\omega_m)^{2}}
\end{eqnarray}
where $P^{0}$ is a positive constant chosen for a continuous
match.

The correlated $GW$ self-energy is given by
\begin{eqnarray}\label{GWwithDC}
\Sigma^{c}_{ij\sigma}({\bf q};i\nu_n) &=&
-\frac{1}{\beta}\sum^{N_{s}(n)}_{m\geq0}\frac{1}{N_k}\sum_{{\cal
R}}\sum_{{\bf k}\in {\rm IBZ}}[G_{ij\sigma}({\bf q}-{\cal R}{\bf
k};i\nu_{n}-i\omega_{m})W^{c}_{ij}({\bf k};i\omega_{m})+
\nonumber\\&& \hspace{96pt}G_{ij\sigma}({\bf q}-{\cal R}{\bf
k};i\nu_{n}+i\omega_{m+1})(W^{c})^{*}_{ij}({\bf
k};i\omega_{m+1})].
\end{eqnarray}
The first term, is large whenever $(m-n)$ around zero, due to
$G(m-n)$, even if the screened interaction is decaying for large
$m$. This means that the upper limit for the $m$-sum should depend
on $n$. We have performed the sum for $m=0,N_{s}(n)$ where
$N_{s}(n)=N_{S}+n$.

The impurity (Anderson Hamiltonian) is solved using the updated
effective ${\cal U}$, {\em not} the bare $U$. To be consistent,
the localized level in the impurity model is updated using
$\varepsilon_d=-\mu=-({\cal U}/2 + \Delta \mu)$. In the
half-filled case $\Delta \mu=0$ (hole doping $\Delta \mu<0$). We
have scaled the bath ${\cal G}^{-1}$ as well as the impurity
self-energy $\Sigma_{i\sigma}$. We have
\begin{eqnarray}
G^{-1}_{ij\sigma}({\bf k};i\nu_n) &=& (i\nu_n + \mu)\delta_{ij} +
h_{ij}({\bf k}) - \Sigma_{ij\sigma}({\bf k};i\nu_n)
\nonumber\\
&=&(i\nu_n + {\cal U}/2+\Delta
\mu-\Sigma_{i\sigma}(i\nu_n))\delta_{ij} + h_{ij}({\bf k}) -
\Sigma^{GW}_{ij\sigma}({\bf k};i\nu_n)\nonumber\\
&=&(i\nu_n +\Delta \mu-[\Sigma_{i\sigma}(i\nu_n)-{\cal
U}/2])\delta_{ij} + h_{ij}({\bf k}) - \Sigma^{GW}_{ij\sigma}({\bf
k};i\nu_n)\nonumber\\
&=&(i\nu_n +\Delta \mu-\tilde
\Sigma_{i\sigma}(i\nu_n))\delta_{ij} + h_{ij}({\bf k}) -
\Sigma^{GW}_{ij\sigma}({\bf k};i\nu_n).
\end{eqnarray}
The GW self-energy includes the double counting term. We also have
\begin{eqnarray}
{\cal G}^{-1}_{i\sigma}(N_s,i\nu_n) &=& (i\nu_n - \varepsilon_d)
-
\sum^{N_s - 1}_{k=1} \frac{V^{2}_{ik\sigma}}{i\nu_n-\varepsilon_{ik\sigma}}\nonumber\\
&=&  (i\nu_n + \mu)  -
\sum^{N_s - 1}_{k=1} \frac{V^{2}_{ik\sigma}}{i\nu_n-\varepsilon_{ik\sigma}}\nonumber\\
&=&(i\nu_n + {\cal U}/2+\Delta \mu)  - \sum^{N_s - 1}_{k=1}
\frac{V^{2}_{ik\sigma}}{i\nu_n-\varepsilon_{ik\sigma}}.
\end{eqnarray}
Thus
\begin{eqnarray}
 \tilde {\cal G}^{-1}_{i\sigma}(N_s,i\nu_n) &=& (i\nu_n + \Delta\mu)  -
\sum^{N_s - 1}_{k=1}
\frac{V^{2}_{ik\sigma}}{i\nu_n-\varepsilon_{ik\sigma}}
\end{eqnarray}
with $ \tilde {\cal G}^{-1}\equiv{\cal G}^{-1}-{\cal U}/2$. The
self-consistency relation:
\begin{eqnarray}
{\cal G}^{-1}_{i\sigma}(i\nu_n)&=& G^{-1}_{i\sigma}(i\nu_n)
+\Sigma_{i\sigma}(i\nu_n)
\nonumber\\
{\cal G}^{-1}_{i\sigma}(i\nu_n)-{\cal U}/2&=&
G^{-1}_{i\sigma}(i\nu_n) +\Sigma_{i\sigma}(i\nu_n)-{\cal U}/2
\nonumber\\
\tilde{\cal G}^{-1}_{i\sigma}(i\nu_n)&=& G^{-1}_{i\sigma}(i\nu_n)
+\tilde\Sigma_{i\sigma}(i\nu_n)
\end{eqnarray}
We note that the Hartre-Fock (impurity) self-energy can be
written as
\begin{eqnarray}
\Sigma^{HF}_{i\sigma} = {\cal U}n_{i-\sigma}.
\end{eqnarray}
In the half-filled case $n_{i-\sigma}=1/2$ for all sites $i$ and
spin-channels, so $\tilde\Sigma_{i\sigma}(i\nu_n)$ is the
impurity self-energy with the static Hatree-Fock part removed.
\\\\
Iterative steps:
\\ \\
1. For each site $i$ in the unit cell and spin-channel $\sigma$,
we have to solve an impurity problem. The  Anderson Hamiltonian,
which is defined by  $\varepsilon_d$,
$\{\epsilon_{ik\sigma},V_{ik\sigma}\}$ and the effective Hubbard
${\cal U}_i$, is solved in order to get the impurity Green's
function $G_{i\sigma}$ and the static response
$\chi_{i}(i\omega_m= 0)$. Using the response we can calculate the
screened interaction for the impurity; $W_{i}(i\omega_m= 0) =
{\cal U}_i + {\cal U}_i\chi_{i}(i\omega_m= 0){\cal U}_i$ as well
as the impurity polarization $P_{i}(i\omega_m= 0) = {\cal
U}^{-1}_i - W^{-1}_{i}(i\omega_m= 0)$.
\\ \\
2. Derive the (scaled ) impurity self-energy from
$\tilde\Sigma_{i\sigma}(i\nu_n)=\tilde{\cal
G}^{-1}_{i\sigma}(i\nu_n)- G^{-1}_{i\sigma}(i\nu_n)$. Here we use
the bath Green's function from the previous iteration. In the
first iteration we have to guess the Anderson (bath) parameters as
well as the bath Green's function.
\\ \\
3. With the impurity self-energy we construct
($\Sigma^{GW}_{ij\sigma}({\bf k};i\nu_n)$ from the previous
iteration which includes the double counting term for $i=j$)
\begin{eqnarray}
G^{-1}_{ij\sigma}({\bf k};i\nu_n)  &=&(i\nu_n +\Delta \mu-\tilde
\Sigma_{i\sigma}(i\nu_n))\delta_{ij} + h_{ij}({\bf k}) -
\Sigma^{GW}_{ij\sigma}({\bf k};i\nu_n).
\end{eqnarray}
Using $G_{ij\sigma}({\bf k};i\nu_n)$ we construct the updated GW
self-energy to be used in the next iteration and then we
calculate the local Green's function $\sum_{{\bf k}}
G_{ii\sigma}({\bf k};i\nu_{n})$ using the impurity self-energy
and the {\em updated} GW self-energy. We also calculate the local
screened interaction $\sum_{{\bf k}}W_{ii}({\bf k};i\omega_m= 0)$.
\\ \\
 4. Update bath
Green's function using $\tilde{\cal G}_{i\sigma}(i\nu_n)=
[1/\sum_{{\bf k}} G_{ii\sigma}({\bf k};i\nu_{n})
+\tilde\Sigma_{i\sigma}(i\nu_n)]^{-1}$ and the effective
interaction using ${\cal U}_{i}= [1/ \sum_{{\bf k}}W_{ii}({\bf
k};i\omega_m= 0) +P_{i}(i\omega_{m}=0)]^{-1}$.
\\ \\
5. Mix old (bath $\tilde{\cal G}$ used in step 2) and new (bath
$\tilde{\cal G}$ from step 4). Same mixing for old effective
interaction ( ${\cal U}$ used in step 1) and new ( ${\cal U}$
from step 4).
\\ \\
6. The mixed bath Green's function $\tilde {\cal G}^{-1}$ is then
fitted ($\tilde{\cal G}^{-1}_{i\sigma}(i\nu)\approx\tilde{\cal
G}^{-1}_{i\sigma}(N_s,i\nu)$) in order to determine the updated
parameters $\{\epsilon_{ik\sigma},V_{ik\sigma}\}$.
\\ \\
7. Now we have a new set of parameters (which defines the impurity
problem) so we go back to step 1. We also have a new bath
$\tilde{\cal G}$ to be used in step 2. At self-consistency the
Green's function obtained from the impurity problem is equal to
local one obtained from $\sum_{{\bf k}} G_{ii\sigma}({\bf
k};i\nu_{n})$ and the impurity screened interaction is identical
to $\sum_{{\bf k}}W_{ii}({\bf k};i\omega_m= 0)$.
\\ \\

\section{Results and discussion}
We use a simple model system  as a test of the feasibility of the
method and therefore consider a one-band Hubbard model. It is
worth to point out that we are mainly interested in how
properties, derived using the DMFT, {\em changes} when GW effects
are incorporated as well as the the stability of the iterative
procedure. At self-consistency we have access to the {\em full}
self-energy and polarization operator as well as ${\cal G}$ and
${\cal U}$. In this work we focus on the PM solution at
half-filling (one electron per site) but not to close to  the
metal-insulator transition. We believe that a careful analysis of
the fictitious temperature and the number of bath sites is not so
crucial when the system is quite far from the metal-insulator
transition. All results presented here will be for four bath-sites
($N_{s}=4$). The system studied  consists of two sites in the unit
cell (denoted 1 and 2) and we impose no constraints on different
sites and spin-channels i.e in the paramagnetic case we will
obtain four identical solutions. If the system initially is in the
metallic PM phase, the system can during the iteratively
procedure, end up stable in the insulator AF phase when
convergency is reached. Such a scenario is of course not possible
if only one site and spin is considered per unit cell. We will
assume that all energies are given in eV.

\subsection{1D chain}

\begin{figure}[H]
\begin{center}
\centerline{
\includegraphics[height=6cm]{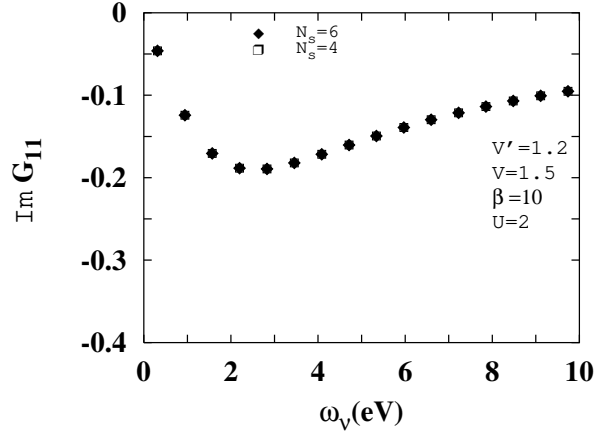}}
\vspace{0.5cm} \caption[]{ Imaginary part of the site-diagonal Green's
 function at the $\Gamma$-point for $N_{s}=4$ and $6$. Parameters used for DMFT+GW:
$N_{k}=33$, $N_{g}=512$, $N_{h}=128$ and $N_{P}=N_{S}=64$.
 } \label{fig:convNs11}
\end{center}
\hfill
\end{figure}

Although a Luttinger liquid we will consider the 1D chain
(bandwidth 4) and we have chosen $U=2$  and $U=14$ as prototypes
for a metal and an insulator respectively. We have checked the
convergency with respect to the number of bath-sites.
% Luttinger liquid: no jump in n_{k} at the Fermi level, no quasiparticles
% only incoherent weight at the Fermi level. Could be metallic.
 In  Fig. \ref{fig:convNs11} the imaginary
part of the on-site lattice Green's function is displayed as a
function of imaginary (odd Matsubara) frequencies corresponding to
the inverse temperature $\beta$. Convergency test with respect to
the number of points in the (1BZ) in addition to the energy-range
parameters ($N_{g}$, $N_{h}$,  $N_{P}$ and $N_{S}$ defined in
paragraph II C) has been performed as well. We will first discuss
a typical metal. Apart from the on-site interaction (short-ranged)
$U$ the present GW approach also contains the off-site
(long-ranged) interactions $V$ and $V'$ (see Eq. \ref{bareV}).
Quite naturally the significance of the GW effects are in some
sense tuned by the magnitude of these off-site interactions. We
have chosen the parameters $V=1.5$ and $V'=1.2$  in the metal case
$U=2$. This choice of parameters are not, at this point, dictated
by any physical grounds. However we believe that parameters chosen
are in a such a range that at least some comprehensive statements
can be made. The difference between using $\beta = 10$ or $\beta =
20$ is very small (the number of Matsubara energy points in the
low temperature case was increased correspondingly) and if not
stated otherwise the inverse temperature is $\beta =10$.

\begin{figure}[H]
\begin{center}
\centerline{
\includegraphics[height=6cm]{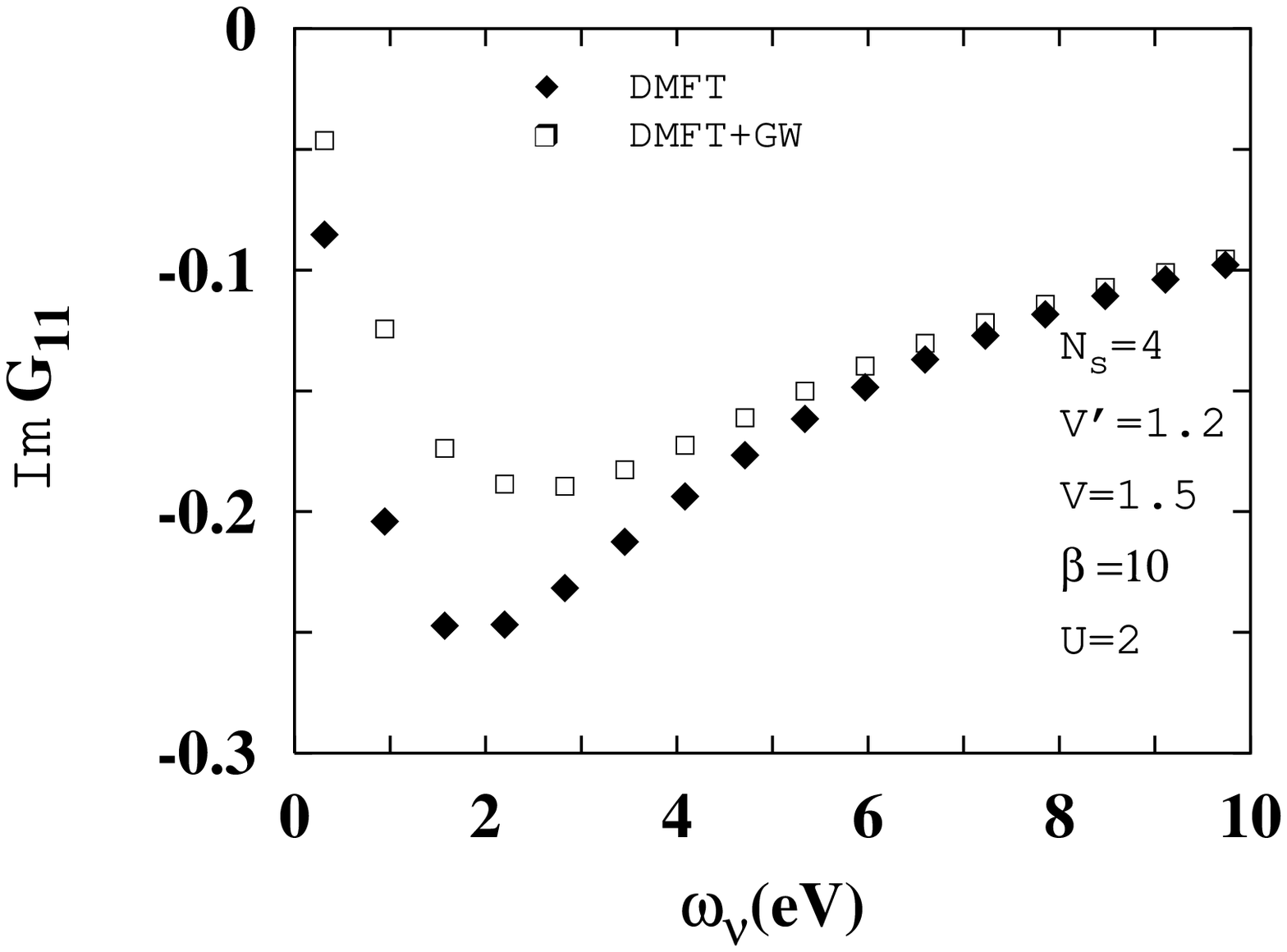}}
\vspace{0.5cm} \caption[]{ Imaginary part of the site-diagonal Green's
 function at the $\Gamma$-point.
Parameters used for DMFT+GW:
$N_{k}=33$, $N_{g}=512$, $N_{h}=128$ and $N_{P}=N_{S}=64$.
 } \label{fig:G11U2}
\end{center}
\hfill
\end{figure}

\begin{figure}[H]
\begin{center}
\centerline{
\includegraphics[height=6cm]{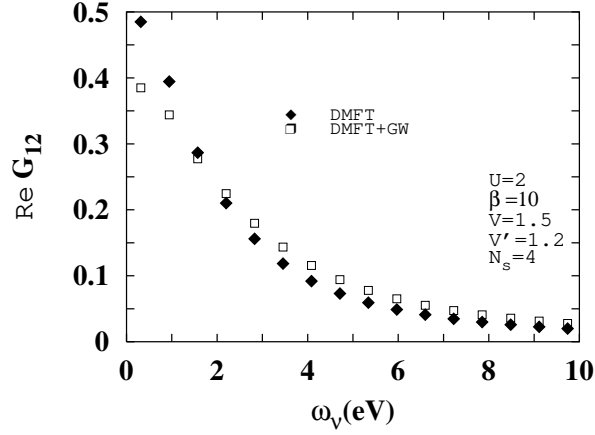}}
\vspace{0.5cm} \caption[]{ Real part of the site-nondiagonal Green's
 function at the $\Gamma$-point. Parameters used for DMFT+GW:
$N_{k}=33$, $N_{g}=512$, $N_{h}=128$ and $N_{P}=N_{S}=64$.
 } \label{fig:G12U2}
\end{center}
\hfill
\end{figure}

In Figs. (\ref{fig:G11U2}-\ref{fig:G12U2}) the $k$-dependent
lattice Green's functions
are shown in the low-energy region. In the DMFT case the total self-energy
is merely composed of
the local impurity ($k$-independent)
self-energy defined in Eq. 14 ($\Sigma^{GW}=0)$.
Obviously the inclusion of the GW self-energy is quite substantial
for small energies. We like to stress that
the total self-energy (Eq. \ref{totsig})
exhibits {\em non-diagonal} site contributions
originating from the GW kernel, influencing the Green's function and
consequently the spectral properties.
 The displayed behaviour of the Green's function has been observed by
several authors
\cite{capone,gw1,gw2,cluster5}. Capone {\em et al.}\cite{capone}
 have found, in the metallic
region, that the inclusion of a cluster DMFT approach will give rise to
a dip in the in the imaginary part of the on-site
Green's function, albeit characteristic of an insulator.

\noindent

\noindent
\noindent

\begin{figure}[H]
\begin{center}
\centerline{
\includegraphics[height=6cm]{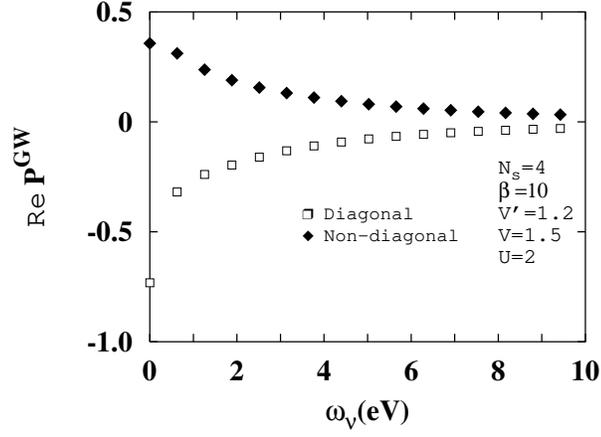}}
\vspace{0.5cm} \caption[]{ Real part of the polarization function
 at the $\Gamma$-point.
The static impurity contribution is $P_{1}(i\omega_m= 0)$ = -0.47.
Parameters used for DMFT+GW:
$N_{k}=33$, $N_{g}=512$, $N_{h}=128$ and $N_{P}=N_{S}=64$.
 } \label{fig:P0U2}
\end{center}
\hfill
\end{figure}

\begin{figure}[H]
\begin{center}
\centerline{
\includegraphics[height=6cm]{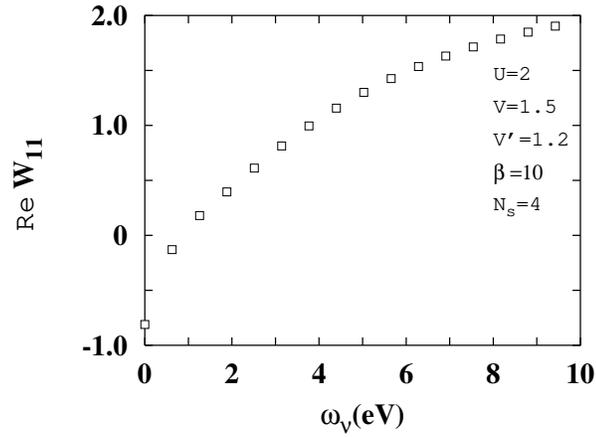}}
\vspace{0.5cm} \caption[]{ Real part of the site-diagonal
(correlated ) screened interaction
 at the $\Gamma$-point. The bare Hubbard $U$ has been subtracted.
The impurity screened
interaction is 0.5 and the effective Hubbard ${\cal U}=0.7$.
Parameters used for DMFT+GW:
$N_{k}=33$, $N_{g}=512$, $N_{h}=128$ and $N_{P}=N_{S}=64$.
 } \label{fig:W11U2}
\end{center}
\hfill
\end{figure}

\begin{figure}[H]
\begin{center}
\centerline{
\includegraphics[height=6cm]{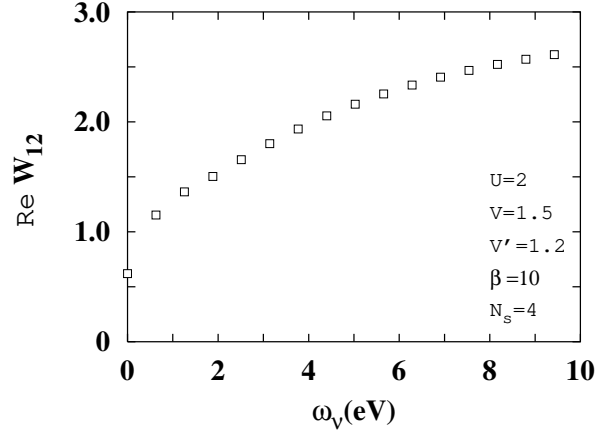}}
\vspace{0.5cm} \caption[]{ Real part of the site-nondiagonal
screened interaction
 at the $\Gamma$-point.
Parameters used for DMFT+GW: $N_{k}=33$, $N_{g}=512$, $N_{h}=128$
and $N_{P}=N_{S}=64$.
 } \label{fig:W12U2}
\end{center}
\hfill
\end{figure}

\begin{figure}[H]
\begin{center}
\centerline{
\includegraphics[height=6cm]{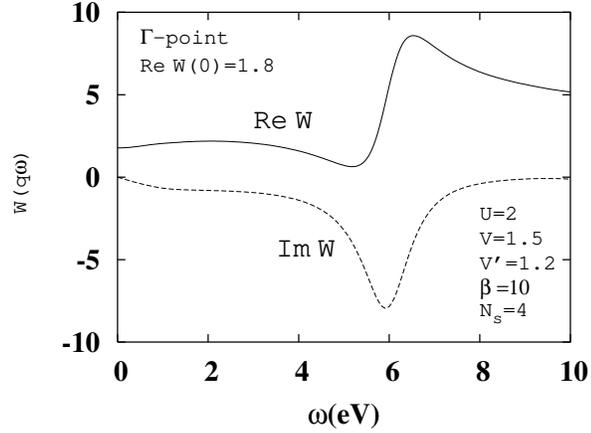}}
\vspace{0.5cm} \caption[]{ Real and imaginary part of
the site-diagonal
screened interaction
 at the $\Gamma$-point.
We used an artificial broadening of 0.5.
Parameters used for DMFT+GW:
$N_{k}=33$, $N_{g}=512$, $N_{h}=128$ and $N_{P}=N_{S}=64$.
 } \label{fig:WwU2}
\end{center}
\hfill
\end{figure}

The GW derived polarization (Eq. \ref{Pij0}) and screened
interaction  (correlated part) (Eq. \ref{Wij}) are displayed in
Figs. (\ref{fig:P0U2}-\ref{fig:W12U2}) as a function of imaginary
(even Matsubara) frequencies. Note that if the {\em full}
polarization in Eq. \ref{fullP} is required one has to correct for
double counting and merely add the static impurity contribution
($P_{1}(i\omega_m= 0)$ = -0.47). For large Matsubara energies  the
diagonal part approaches $2V'$ and the non-diagonal part $2V$ as
can be derived using Eq. (\ref{hopp}),  which is numerically
confirmed. In Fig. (\ref{fig:WwU2}) we show the screened
interaction at the $\Gamma$- point along the real axis using the
Pade approximation. We observe that the static value ${\rm
Re}W(0)$ is merely a constant below the main excitation peak and
slightly larger in the case of non-diagonal screening. However it
is well-known that in the RPA the screening is overestimated at
short distances. From a physical point of view this fact is easily
understandable: a positive hole is surrounded or screened by a too
tightly drawn electron cloud, due to the fact that exchange and
correlation effects are neglected among the screening electrons.

\noindent

\noindent
\noindent

\begin{figure}[H]
\begin{center}
\centerline{
\includegraphics[height=6cm]{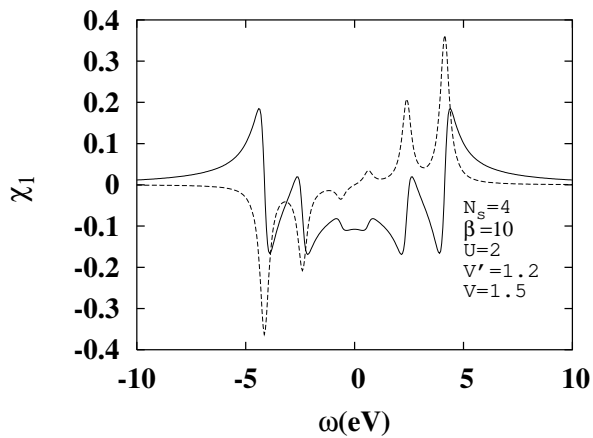}}
\vspace{0.5cm} \caption[]{Real (solid line) and imaginary (
dashed line) part of the impurity response function (site 1) defined in
Eq. \ref{chargeres} for the effective impurity.
Parameters used for DMFT+GW:
$N_{k}=33$, $N_{g}=512$, $N_{h}=128$ and $N_{P}=N_{S}=64$.
 } \label{fig:XwU2}
\end{center}
\hfill
\end{figure}

The self-consistent values of the impurity screened potential
and the effective Hubbard on-site interaction
(both defined in Eq. \ref{Wimpur})
were found to be  $W(i\omega_{m}=0)=0.5$  and ${\cal U}=0.7$  respectively.
Thus at self-consistency,
the {\em effective} impurity problem offers an on-site interaction
which is more than a factor of two smaller than the bare $U=2$.
For  illustration we show the charge {\em impurity}
 response function
along the real axis in
Fig. \ref{fig:XwU2} derived using the effective Hubbard ${\cal U}=0.7$.

\noindent
%\noindent
%\begin{figure}[H]
%\begin{center}
%\centerline{
%\includegraphics[height=6cm]{Ww12U2.ps}}
%\vspace{0.5cm} \caption[]{ Real and imaginary part of
%the site-nondiagonal
%screened interaction
% at the $\Gamma$-point. We used an artificial broadening of 0.5.
%Parameters used for DMFT+GW:
%$N_{k}=33$, $N_{g}=512$, $N_{h}=128$ and $N_{P}=N_{S}=64$.
% } \label{fig:Ww12U2}
%\end{center}
%\hfill
%\end{figure}

\noindent

As discussed previously,
 in the single site DMFT case the solution to the impurity model is extracted
 using the bare Hubbard $U=2$, however
in the DMFT+GW scenario the impurity model is solved with the
effective (weaker) interaction ${\cal U}$. The magnitude of the impurity
self-energy scales with the size of the on-site interaction making it somewhat
cumbersome to compare different impurity self-energies obtained
with different on-site interaction strengths. However the quantity one really
should compare is the total self-energy entering the theory i.e
the $U=2$ impurity DMFT single site self-energy should be compared with the
full self-energy in Eq. (\ref{totsig}). In this work a critically comparison
will not be done, we briefly
discuss spectral properties, which however strongly
depends on the self-energy.

Prior to the discussion about spectral properties we intend to make some
statements about the derived self-energies.
In all the cases studied we have observed the characteristics of a metal
or an insulator\cite{rev2}: ${\rm Im}\Sigma(i\omega)<0$
increases linearly or
diverges when $\omega \rightarrow 0^{+}$ respectively.
Regarding the magnitude of the GW self-energy it depends strongly on
the $k$-point,
but in general the non-diagonal
${\rm Re} \Sigma^{GW}_{12}$ is quite large and ${\rm Im} \Sigma^{GW}_{11}$
is smaller (in comparison with relevant quantities).

\begin{figure}[H]
\begin{center}
\centerline{
\includegraphics[height=6cm]{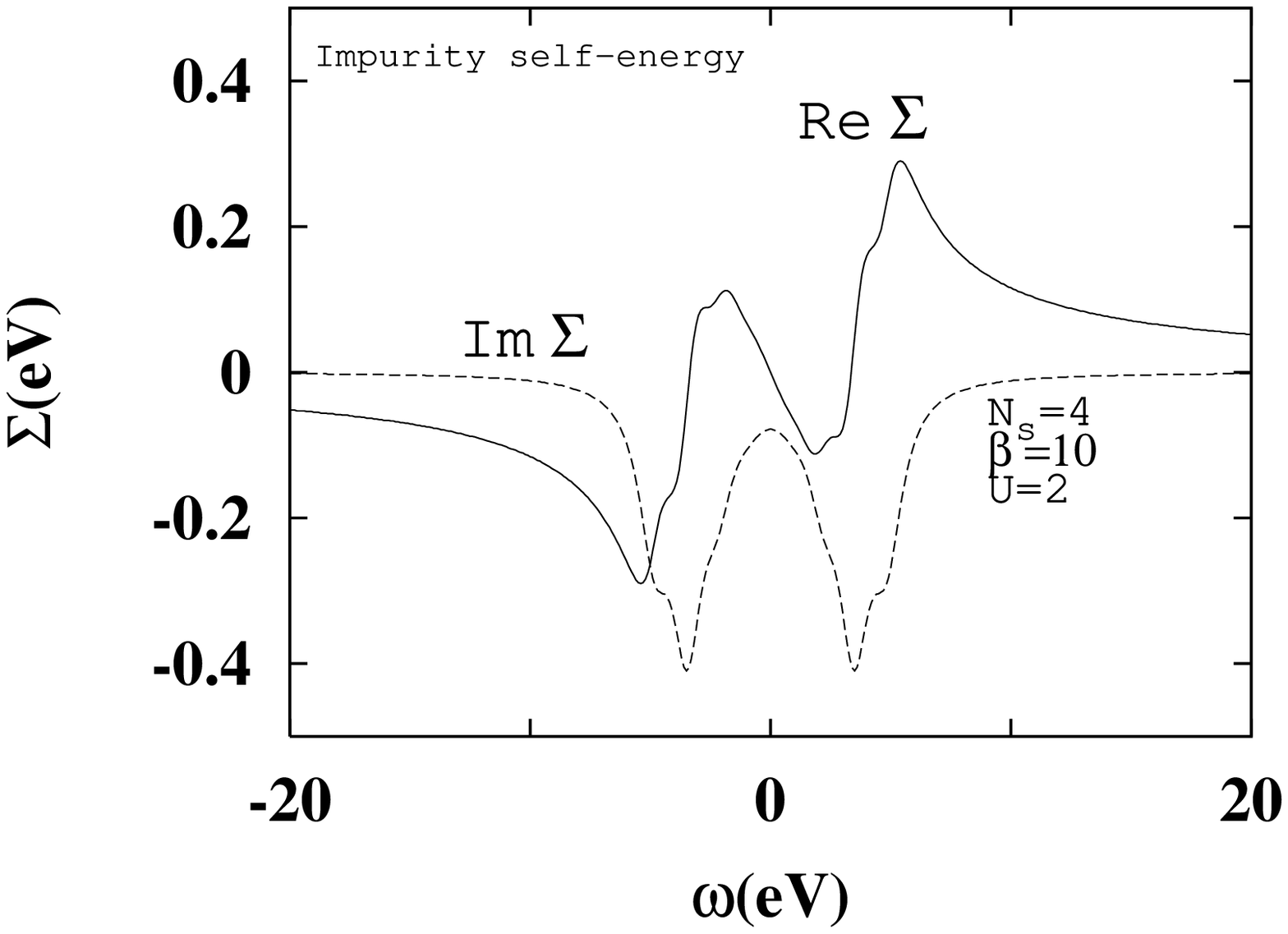}}
\vspace{0.5cm} \caption[]{ Real and imaginary part of the
impurity self-energy (site 1 and spin up) for a typical metal.
We used an artificial broadening of 0.75.
Parameters used for DMFT:
$N_{k}=33$ and $N_{g}=512$.
 } \label{fig:SwImpU2}
\end{center}
\hfill
\end{figure}

\begin{figure}[H]
\begin{center}
\centerline{
\includegraphics[height=6cm]{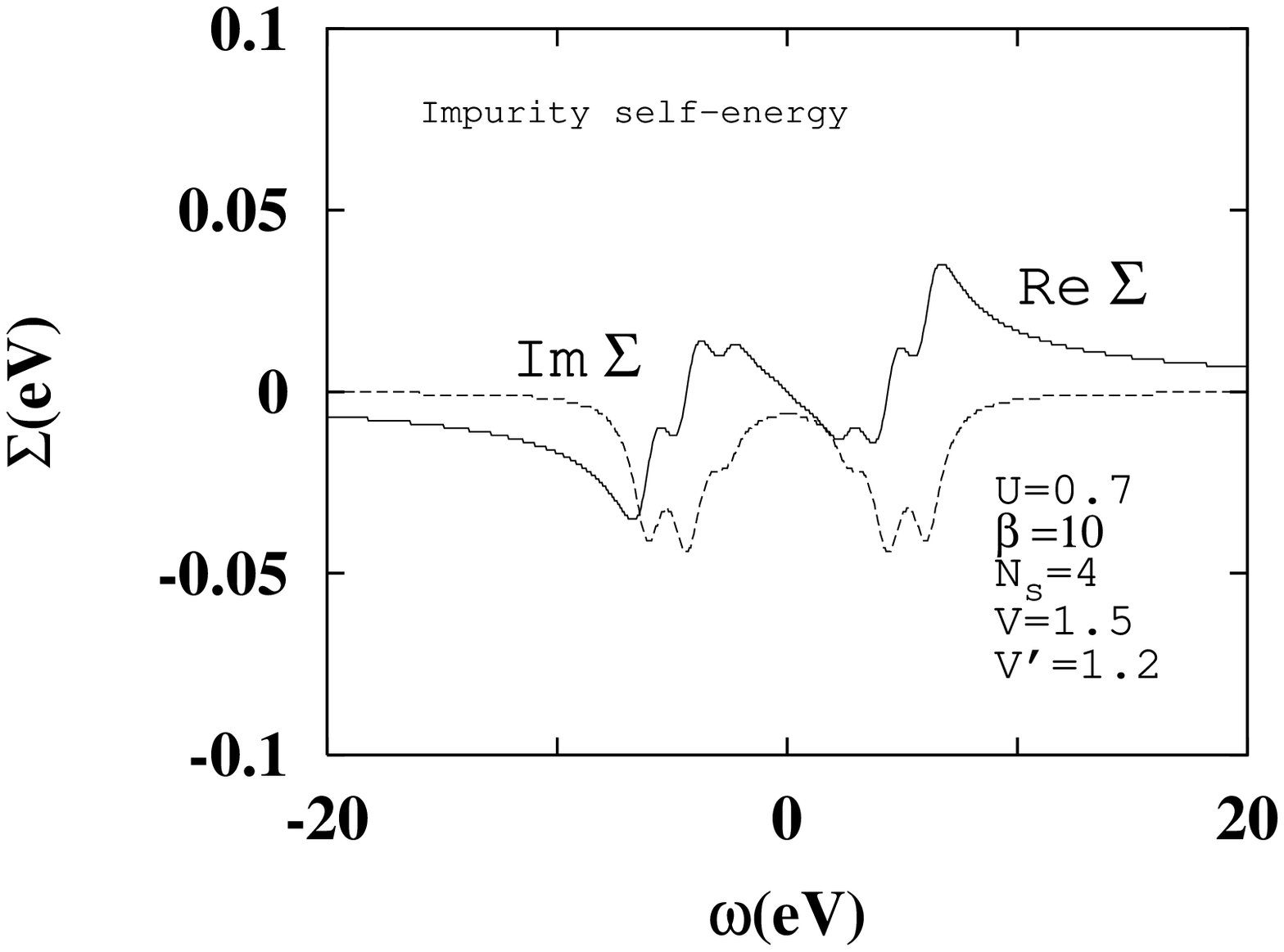}}
\vspace{0.5cm} \caption[]{ Real and imaginary part of the
impurity self-energy (site 1 and spin up) for a typical metal.
We used an artificial broadening of 0.75.
 Parameters used for DMFT+GW:
$N_{k}=33$, $N_{g}=512$, $N_{h}=128$ and $N_{P}=N_{S}=64$.
 } \label{fig:SwImpU07}
\end{center}
\hfill
\end{figure}

\begin{figure}[H]
\begin{center}
\centerline{
\includegraphics[height=6cm]{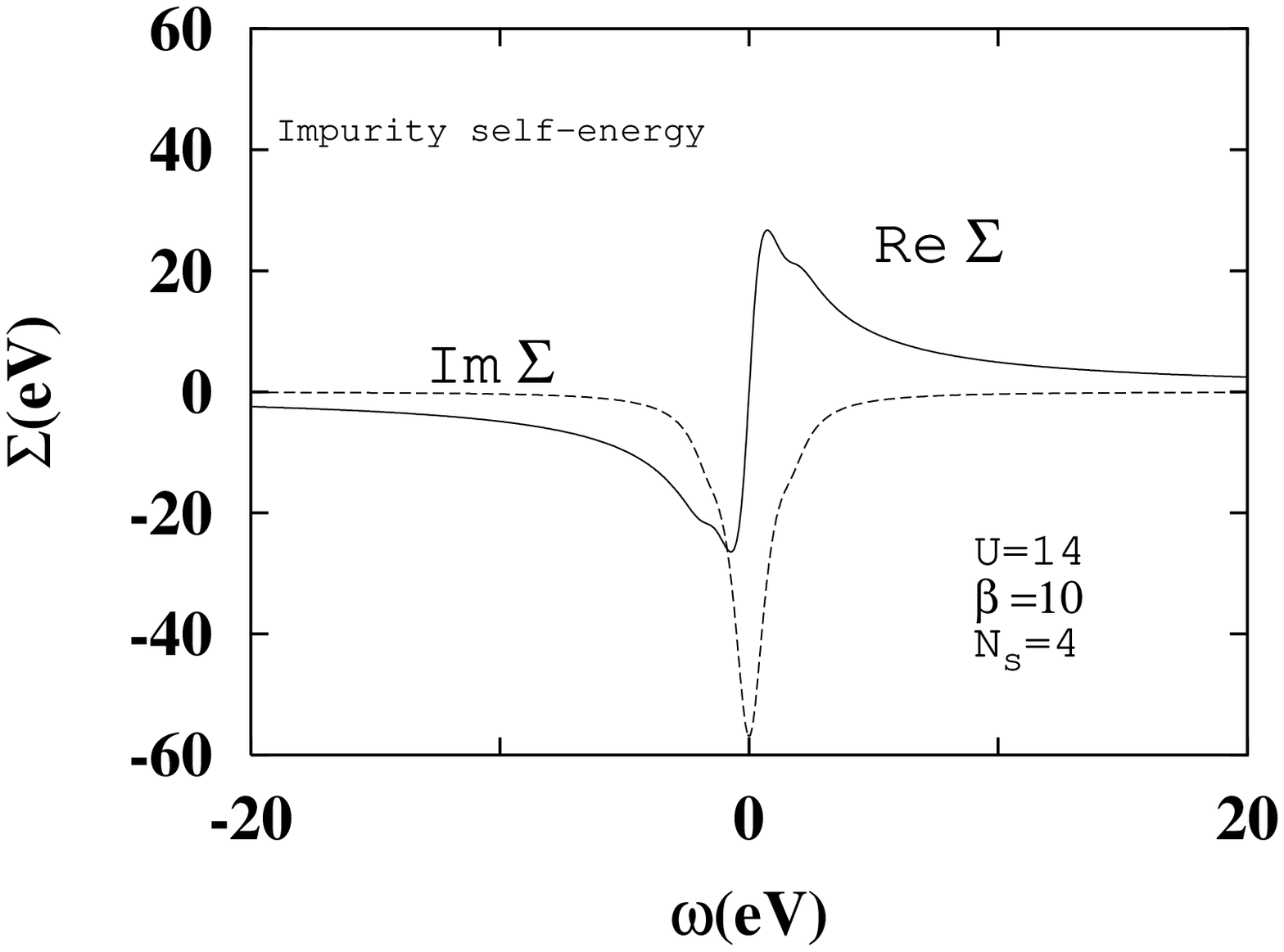}}
\vspace{0.5cm} \caption[]{ Real and imaginary part of the
impurity self-energy (site 1 and spin up) for a typical insulator.
We used an artificial broadening of 0.75.
Parameters used for DMFT:
$N_{k}=33$ and $N_{g}=512$.
 } \label{fig:SwImpU14}
\end{center}
\hfill
\end{figure}

\begin{figure}[H]
\begin{center}
\centerline{
\includegraphics[height=6cm]{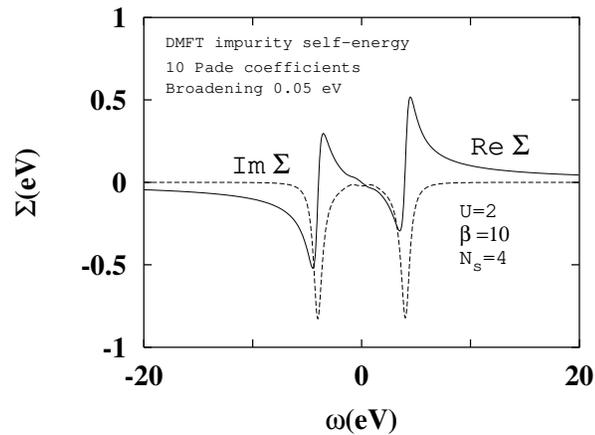}}
\vspace{0.5cm} \caption[]{ Real and imaginary part of the
impurity self-energy (site 1 and spin up) for a typical metal.
Parameters used for DMFT:
$N_{k}=33$ and $N_{g}=512$.
 } \label{fig:SwImp-padeU2}
\end{center}
\hfill
\end{figure}

It is a delicate matter to extract real frequency dynamical information
from imaginary axis data. The commonly used
quantum Monte Carlo impurity solver uses
maximum entropy based methods\cite{jarell}. In the present work,
adopting the Lanczos routine for solving the impurity
problem,  we used the the Pade approximation when performing
the analytical continuation.
For example in order to obtain the GW self-energy and spectral functions
we must do an analytical continuation from the imaginary axis
($i\omega \rightarrow (\omega + i\delta)$).
However,
the impurity self-energy on the real axis can be extracted using the
self-consistency relation in Eq. \ref{scg}. The corresponding
results are
 shown in
Figs. (\ref{fig:SwImpU2}, \ref{fig:SwImpU14}) (DMFT) and Fig.
\ref{fig:SwImpU07} (DMFT+GW).
 In the DMFT+GW scenario the impurity
is solved with an effective Hubbard interaction
${\cal U}= 0.7$, reflected in a substantial reduction in the
magnitude. As a comparison with  Fig. (\ref{fig:SwImpU2}) the self-energy
derived using the
Pade approximation is displayed in Fig. \ref{fig:SwImp-padeU2}.

In the metallic case the self-energy exhibits the Fermi-liquid
behaviour: ${\rm Re} \Sigma(\omega)\sim (1-1/Z)\omega$ (or
equivalently ${\rm Im}\Sigma(i\omega) \sim (1-1/Z)\omega$) and
${\rm Im} \Sigma(\omega)\sim -\omega^{2}$  for $\omega$ close to
zero where
\begin{equation}
Z=(1-{\frac{\partial {\rm Re}
\Sigma(\omega)}{\partial \omega)}\mid_{\omega=0}})^{-1}
\end{equation}
denotes the quasiparticle renormalization factor.
In the insulator case the slope of ${\rm Re} \Sigma(\omega)$ changes sign
 (${\rm Re} \Sigma(\omega)\rightarrow 1/\omega$ for
$\omega \rightarrow 0$)
and ${\rm Im} \Sigma(\omega)$ is peaked at the chemical potential
and zero in the gap\cite{kramer} as evident from Fig. (\ref{fig:SwImpU14}).

\noindent
\noindent

\noindent

\noindent

\begin{figure}[H]
\begin{center}
\centerline{
\includegraphics[height=6cm]{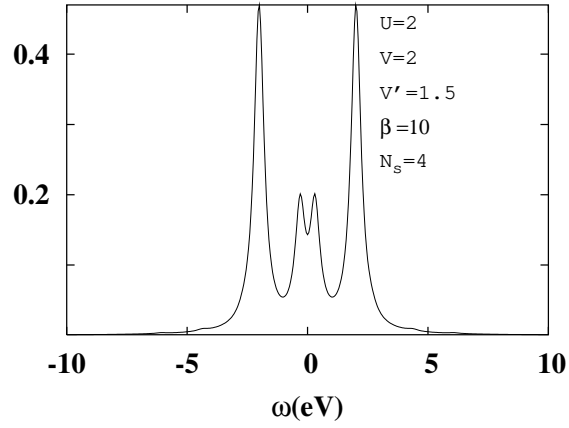}}
\vspace{0.5cm} \caption[]{Local density of states for site 1 and spin up.
The chemical potential
corresponds to energy zero. We introduced an artificial broadening of 0.25.
 Parameters used for DMFT+GW:
$N_{k}=33$, $N_{g}=512$, $N_{h}=128$ and $N_{P}=N_{S}=64$.
 } \label{fig:LDOSU2}
\end{center}
\hfill
\end{figure}

We next discuss spectral properties. In order to achieve the local
density of states (LDOS) we have solved the impurity model on the
real axis and then extracted ${\rm Im} G_{i\sigma}(\omega)$. As
can be seen in Fig. \ref{fig:LDOSU2}, the LDOS is symmetric
(half-filling $n=1$) and shows the typical Fermi metallic
characteristics; a quasiparticle peak surrounding the two Hubbard
bands\cite{rev2}.

\noindent

\begin{figure}[H]
\begin{center}
\centerline{
\includegraphics[height=6cm]{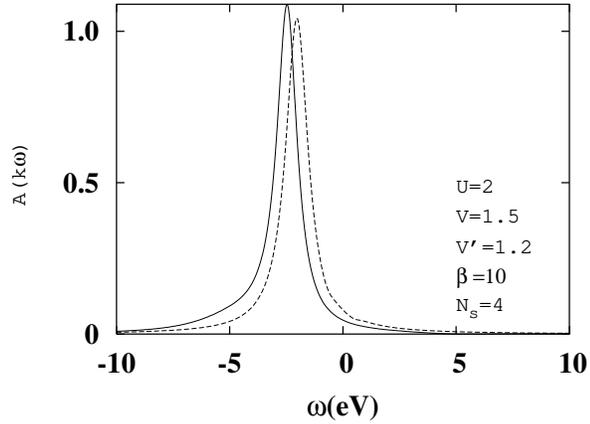}}
\vspace{0.5cm} \caption[]{Spectral function at the $\Gamma$-point.
The chemical potential corresponds to energy zero. The solid line corresponds to the DMFT+GW case and the dashed line to the DMFT case.
%We used 5 Pade coefficients and an artificial broadening of 0.5.
We used an artificial broadening of 0.5.
 Parameters used:
$N_{k}=33$, $N_{g}=512$, $N_{h}=128$ and $N_{P}=N_{S}=64$.
 } \label{fig:AkwU2}
\end{center}
\hfill
\end{figure}

\begin{figure}[H]
\begin{center}
\centerline{
\includegraphics[height=6cm]{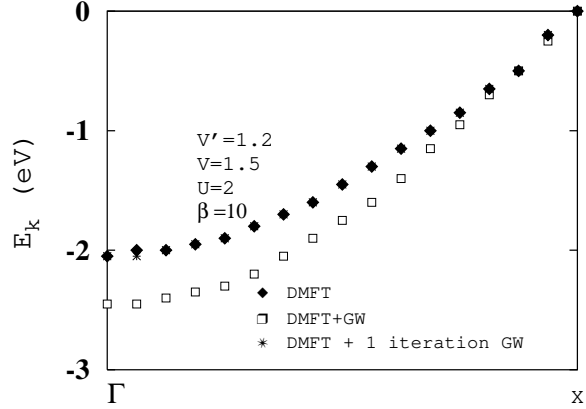}}
\vspace{0.5cm} \caption[]{ Quasiparticle dispersion in the $\Gamma$-$X$
direction.
Parameters used:
$N_{k}=33$, $N_{g}=512$, $N_{h}=128$ and $N_{P}=N_{S}=64$.
 } \label{fig:1Ddisp}
\end{center}
\hfill
\end{figure}

With aid of the Pade approximation and Eq.  \ref{akw}
 we calculate the spectral functions.
The zone-center spectral function is
visualized in Fig. (\ref{fig:AkwU2}).
 A significant change of the quasiparticle peak
position is clearly
seen at the $\Gamma$-point, where the downward shift is around 0.4.
The corresponding dispersion in the $\Gamma$-$X$ direction is displayed in
Fig. \ref{fig:1Ddisp}.
Interestingly  when only {\em one} iteration with the GW  kernel is performed
on top of a self-consistent DMFT calculation, the dispersion
essentially coincide with the
DMFT one.

\noindent
\noindent
\begin{figure}[H]
\begin{center}
\centerline{
\includegraphics[height=6cm]{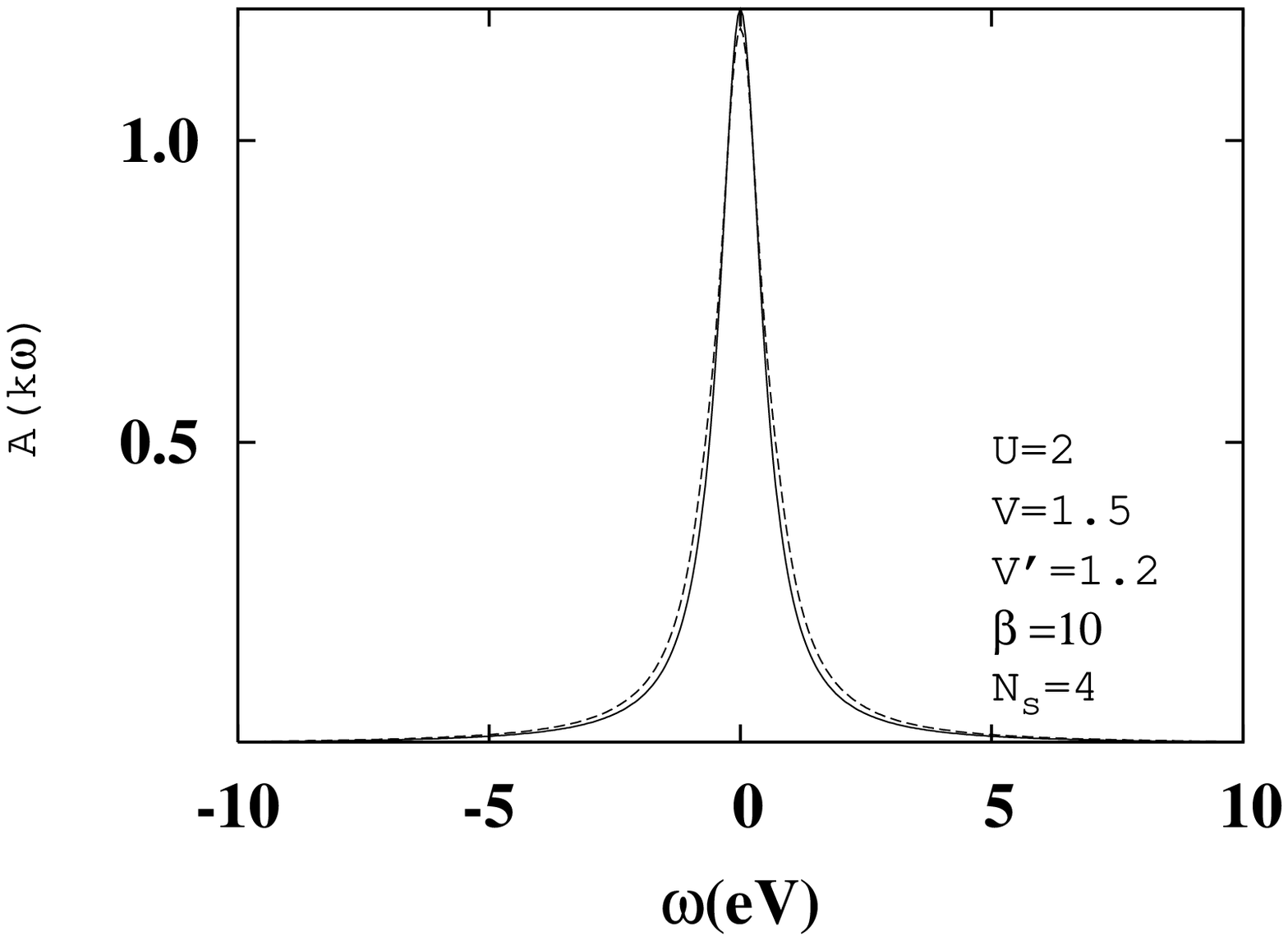}}
\vspace{0.5cm} \caption[]{Spectral function at the $X$-point.
The chemical potential corresponds to energy zero. The solid line corresponds to the DMFT+GW case and the dashed line to the DMFT case.
We used an artificial broadening of 0.5.
%We used 5 Pade coefficients and an artificial broadening of 0.5.
 Parameters used:
$N_{k}=33$, $N_{g}=512$, $N_{h}=128$ and $N_{P}=N_{S}=64$.
 } \label{fig:AkwU2X}
\end{center}
\hfill
\end{figure}
\noindent

\begin{figure}[H]
\begin{center}
\centerline{
\includegraphics[height=6cm]{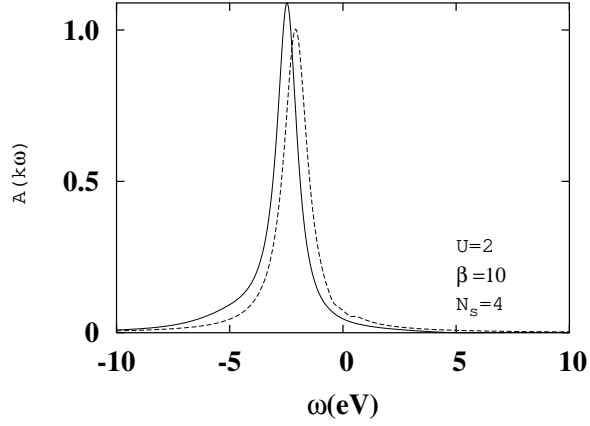}}
\vspace{0.5cm} \caption[]{Spectral function (DMFT+GW) at the $\Gamma$-point .
The chemical potential corresponds to energy zero.
The solid line corresponds to $V=1.5, V'=1.2$ and the dashed line to
$V=V'=0$.
We used an artificial broadening of 0.5.
%We used 5 Pade coefficients and an artificial broadening of 0.5.
 Parameters used:
$N_{k}=33$, $N_{g}=512$, $N_{h}=128$ and $N_{P}=N_{S}=64$.
 } \label{fig:AkwU2VV}
\end{center}
\hfill
\end{figure}

\begin{figure}[H]
\begin{center}
\centerline{
\includegraphics[height=6cm]{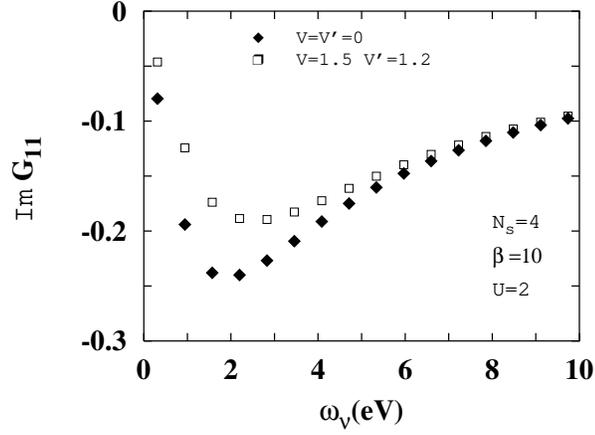}}
\vspace{0.5cm} \caption[]{ Imaginary part of the site-diagonal Green's
 function (DMFT+GW) at the $\Gamma$-point.
Parameters used for DMFT+GW:
$N_{k}=33$, $N_{g}=512$, $N_{h}=128$ and $N_{P}=N_{S}=64$.
 } \label{fig:G11U2VV}
\end{center}
\hfill
\end{figure}

To obtain a notable effect with the GW kernel, one  has in general
to include the long-range part of the bare Coulomb potential and
consider nearest ($V$) and next nearest neighbors interaction
($V'$). For example the parameter-set $V=V'=0$ gives a
quasiparticle peak position shifted only by 0.1 compared to the
DMFT situation (the shift is 0.4 with  $V=1.5, V'=1.2$) which is
realized from Fig. \ref{fig:AkwU2VV}. If one compare the lattice
Green's function in Fig. \ref{fig:G11U2} and  Fig.
\ref{fig:G11U2VV}, it is obvious that DMFT and DMFT+GW with the
long-range part excluded ($V=V'=0$) are quite similar.

\noindent

\noindent

\begin{figure}[H]
\begin{center}
\centerline{
\includegraphics[height=6cm]{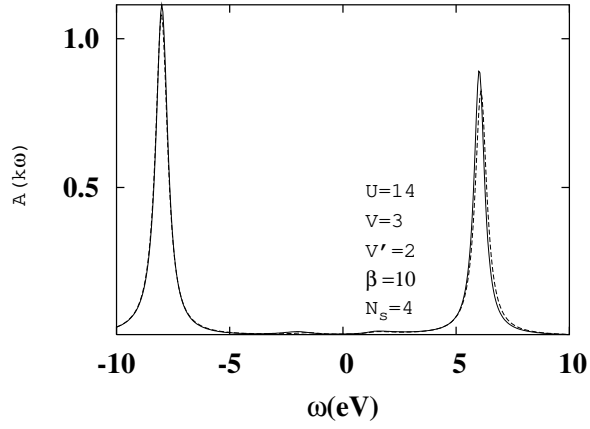}}
\vspace{0.5cm} \caption[]{Spectral function at the $\Gamma$-point.
The chemical potential corresponds to energy zero. The solid line corresponds to the DMFT+GW case and the dashed line to the DMFT case.
We used an artificial broadening of 0.5.
%We used 8 Pade coefficients and an artificial broadening of 0.5.
 Parameters used:
$N_{k}=33$, $N_{g}=512$, $N_{h}=128$ and $N_{P}=N_{S}=64$.
 } \label{fig:AkwU14}
\end{center}
\hfill
\end{figure}

Next we briefly consider a 1D insulator using the parameter-set
$U=14$, $V=3$ and $V'=2$. In contrast to the metal case the
screening is less effective giving the self-consistent impurity
screened interaction to be 13.7 and the effective Hubbard ${\cal
U}=13.9$. The similarities between the screened and bare
interactions indicate that the off-site hopping parameters $V$ and
$V'$ are too small to give rise to a notable effect, which is
indeed confirmed by the
 spectral function shown in
Fig. \ref{fig:AkwU14}.
Furthermore it is worth to note that in the strong insulator case
the imaginary part of the (site-diagonal)
impurity self-energy is diverging for small $i\omega$
 ($\Sigma(i\omega) \rightarrow 1/i\omega$), making at least the significance
of the diagonal GW self-energy negligible. However non-diagonal GW
contributions can influence the spectral functions.

%\noindent
%\begin{figure}[H]
%\begin{center}
%\centerline{
%\includegraphics[height=6cm]{LDOSU14.ps}}
%\vspace{0.5cm} \caption[]{Local density of states for site 1 and spin up.
%The chemical potential
%corresponds to energy zero. We introduced an artificial broadening of 0.25.
% Parameters used for DMFT+GW:
%$N_{k}=33$, $N_{g}=512$ and $N_{h}=N_{P}=N_{S}=128$.
% } \label{fig:LDOSU14}
%\end{center}
%\hfill
%\end{figure}
\noindent
\noindent

\subsection{2D square lattice}
\begin{figure}[H]
\begin{center}
\centerline{
\includegraphics[height=6cm]{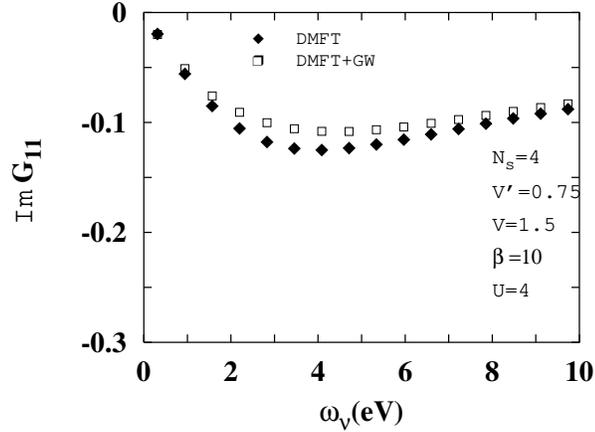}}
\vspace{0.5cm} \caption[]{ Imaginary part of the site-diagonal Green's
 function at the $\Gamma$-point.
Parameters used for DMFT+GW:
$N_{k}=169$, $N_{g}=512$, $N_{h}=128$ and $N_{P}=N_{S}=64$.
 } \label{fig:2DG11U4}
\end{center}
\hfill
\end{figure}

\begin{figure}[H]
\begin{center}
\centerline{
\includegraphics[height=6cm]{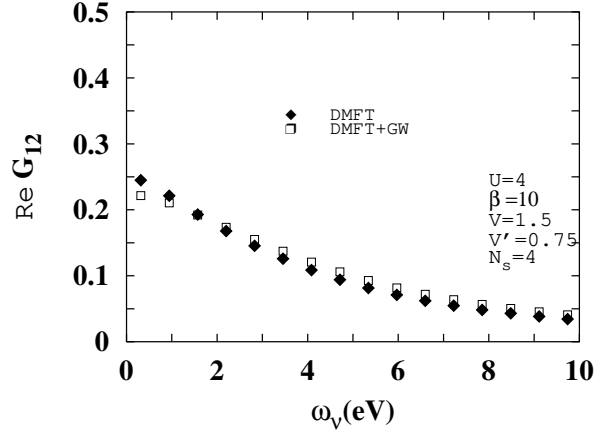}}
\vspace{0.5cm} \caption[]{ Real part of the site-nondiagonal Green's
 function at the $\Gamma$-point.
Parameters used for DMFT+GW:
$N_{k}=169$, $N_{g}=512$, $N_{h}=128$ and $N_{P}=N_{S}=64$.
 } \label{fig:2DG12U4}
\end{center}
\hfill
\end{figure}

The bandwidth of the 2D square lattice is 8 and
we have chosen $U=4$ and $U=18$ as prototypes for a metal and
an insulator respectively. As in 1D $N_{s}=4$ is sufficient.
We will first discuss the metal case. The reasoning and organization follows
closely the setup in the previous section.
We have chosen
the parameters $V=1.5$ and $V'=0.75$  in the metal case $U=4$.
In Figs. (\ref{fig:2DG11U4}-\ref{fig:2DG12U4}) the lattice Green's function
are shown

\noindent

\noindent
\begin{figure}[H]
\begin{center}
\centerline{
\includegraphics[height=6cm]{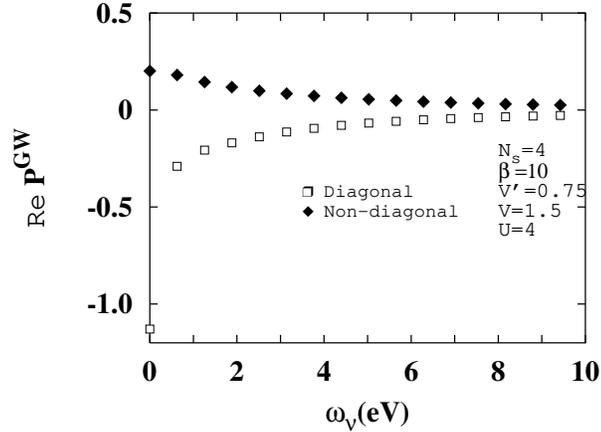}}
\vspace{0.5cm} \caption[]{ Real part of the polarization function
 at the $\Gamma$-point.
The static impurity contribution is $P_{1}(i\omega_m= 0)$ = -1.1.
Parameters used for DMFT+GW:
$N_{k}=169$, $N_{g}=512$, $N_{h}=128$ and $N_{P}=N_{S}=64$.
 } \label{fig:P0U4}
\end{center}
\hfill
\end{figure}
\begin{figure}[H]
\begin{center}
\centerline{
\includegraphics[height=6cm]{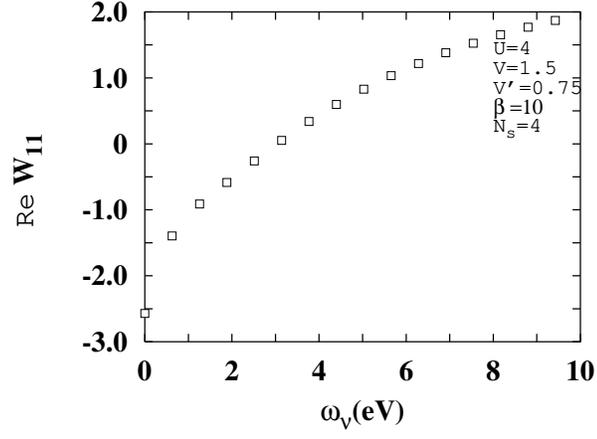}}
\vspace{0.5cm} \caption[]{ Real part of the site-diagonal
(correlated) screened interaction
 at the $\Gamma$-point.  The bare Hubbard $U$ has been subtracted.
Parameters used for DMFT+GW:
$N_{k}=169$, $N_{g}=512$, $N_{h}=128$ and $N_{P}=N_{S}=64$.
 } \label{fig:W11U4}
\end{center}
\hfill
\end{figure}
\begin{figure}[H]
\begin{center}
\centerline{
\includegraphics[height=6cm]{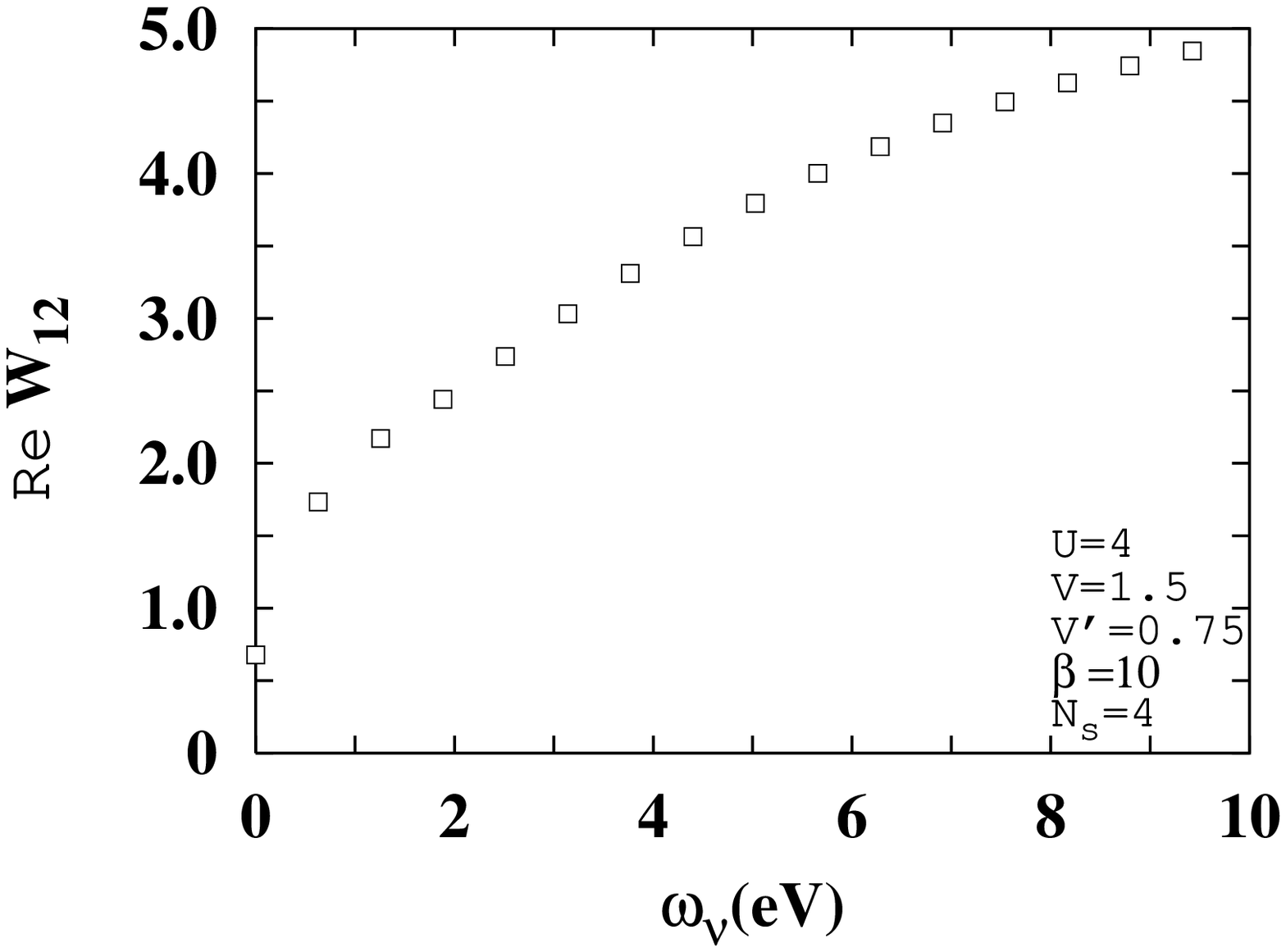}}
\vspace{0.5cm} \caption[]{ Real part of the site-nondiagonal
screened interaction
 at the $\Gamma$-point.
Parameters used for DMFT+GW:
$N_{k}=169$, $N_{g}=512$, $N_{h}=128$ and $N_{P}=N_{S}=64$.
 } \label{fig:W12U4}
\end{center}
\hfill
\end{figure}
\begin{figure}[H]
\begin{center}
\centerline{
\includegraphics[height=6cm]{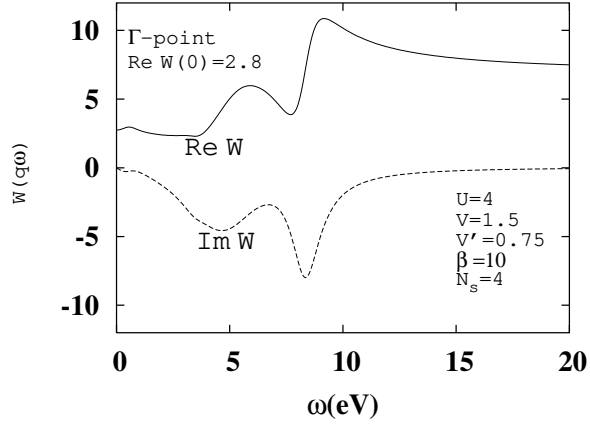}}
\vspace{0.5cm} \caption[]{ Real and imaginary part of
the site-diagonal
screened interaction
 at the $\Gamma$-point.
We used an artificial broadening of 0.5.
Parameters used for DMFT+GW:
$N_{k}=169$, $N_{g}=512$, $N_{h}=128$ and $N_{P}=N_{S}=64$.
 } \label{fig:WwU4}
\end{center}
\hfill
\end{figure}

whereas the corresponding polarization and screened
interaction are displayed
in Figs. (\ref{fig:P0U4}-\ref{fig:W12U4}).
At self-consistency
the impurity screened
interaction is 0.7 and the effective Hubbard ${\cal U}=2.4$,
strongly reduced compared to the bare values. We stress that the amount of
screening that are taking place is in general
dependent of the non-locality parameters
$V$ and $V'$, which in this work is chosen arbitrarily.
In 2D the large $i\omega$ limit is numerically satisfied: the diagonal part
approaches $4V'$ and the
non-diagonal part $4V$ respectively.
It is worth mention that the overall magnitude of the polarization function $P^{GW}(i\omega)$
decreases
for increasing $U$.
As a consequence,
the overall magnitude of the correlated part of the screened interaction
$W^{c}(i\omega)$ increases for increasing $U$.
As a comparison to the 1D case, the screened interaction on real axis
at the $\Gamma$-point is shown in Fig. \ref{fig:WwU4}.

\noindent

\noindent
\noindent

\noindent

\begin{figure}[H]
\begin{center}
\centerline{
\includegraphics[height=6cm]{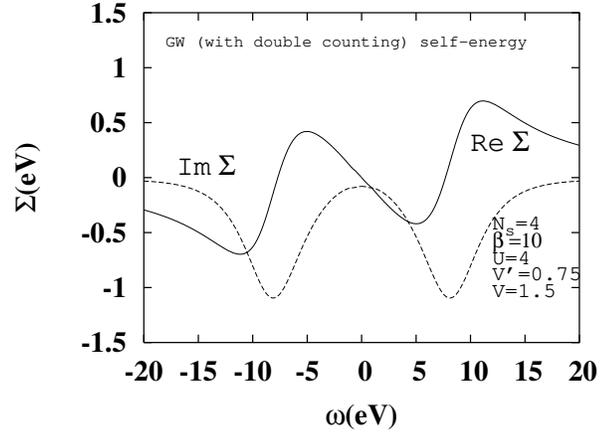}}
\vspace{0.5cm} \caption[]{ Real and imaginary part of the
GW self-energy including the double
counting term (site 1 and spin up) for a typical metal.
We used an artificial broadening of 0.5.
%We used an artificial broadening of 0.5 and 50 Pade coefficients.
 Parameters used for DMFT+GW:
$N_{k}=33$, $N_{g}=512$, $N_{h}=128$ and $N_{P}=N_{S}=64$.
 } \label{fig:SwGWU4}
\end{center}
\hfill
\end{figure}

As an illustration we display  in Fig. (\ref{fig:SwGWU4})
the $GW$ self-energy, which clearly exhibits  Fermi-liquid characteristics,
derived using Eq. \ref{GWwithDC}.

\noindent

\begin{figure}[H]
\begin{center}
\centerline{
\includegraphics[height=6cm]{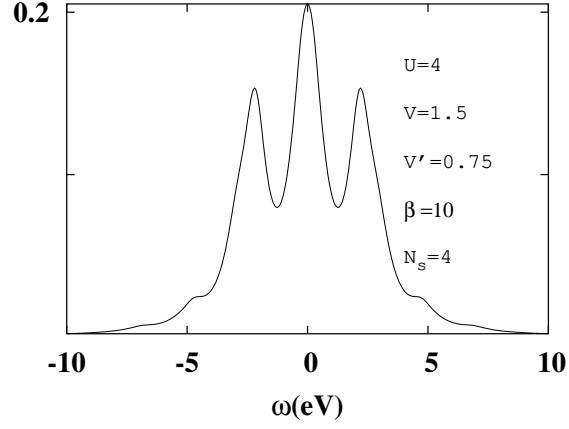}}
\vspace{0.5cm} \caption[]{Local density of states for site 1 and spin up.
The chemical potential
corresponds to energy zero. We introduced an artificial broadening of 0.25.
 Parameters used for DMFT+GW:
$N_{k}=169$, $N_{g}=512$, $N_{h}=128$ and $N_{P}=N_{S}=64$.
 } \label{fig:2DLDOSU4}
\end{center}
\hfill
\end{figure}

\begin{figure}[H]
\begin{center}
\centerline{
\includegraphics[height=6cm]{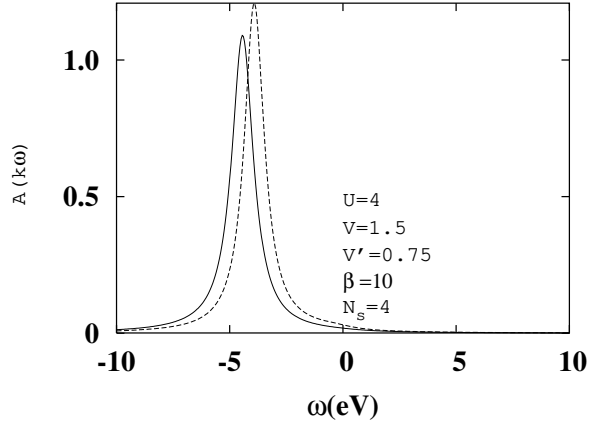}}
\vspace{0.5cm} \caption[]{
Spectral function at the $\Gamma$-point.
The chemical potential corresponds to energy zero.
The solid line corresponds to the DMFT+GW case and the dashed line to
the DMFT case.
We used an artificial broadening of 0.5.
%We used 2 Pade coefficients and an artificial broadening of 0.5.
Parameters used:
$N_{k}=169$, $N_{g}=512$, $N_{h}=128$ and $N_{P}=N_{S}=64$. }
\label{fig:2DAkwU4}
\end{center}
\hfill
\end{figure}

The metallic LDOS and a typical quasiparticle spectral function
are shown in Figs. (\ref{fig:2DLDOSU4}-\ref{fig:2DAkwU4}). The
downward shift of the quasiparticle position is consistent with
the
 scenario observed in the 1D case.

\noindent

\begin{figure}[H]
\centering
\includegraphics[height=6cm]{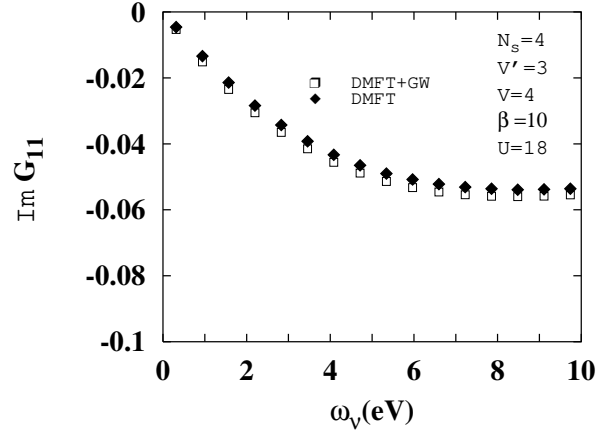}
\caption[]{ Imaginary part of the site-diagonal Green's
 function at the $\Gamma$-point.
Parameters used for DMFT+GW:
$N_{k}=169$, $N_{g}=512$, $N_{h}=128$ and $N_{P}=N_{S}=64$.
 }
\label{fig:2DG11U18}
\end{figure}

\begin{figure}[H]
\begin{center}
\centerline{
\includegraphics[height=6cm]{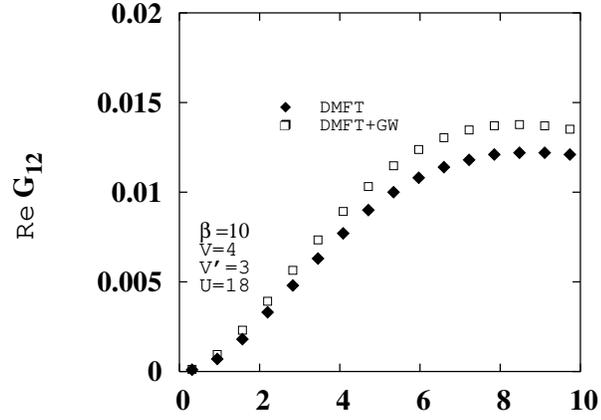}}
\vspace{0.5cm} \caption[]{ Real part of the site-nondiagonal Green's
 function at the $\Gamma$-point.
Parameters used for DMFT+GW:
$N_{k}=169$, $N_{g}=512$, $N_{h}=128$ and $N_{P}=N_{S}=64$.
 } \label{fig:2DG12U18}
\end{center}
\hfill
\end{figure}

Let us finally consider the strong insulator case $U=18$.
We have chosen the parameters $V=4$ and $V'=3$  which can be considered as
a substantial  off-site interaction, however there exists no
large difference in
the DMFT Green's function compared to the DMFT+GW one (see Figs.
(\ref{fig:2DG11U18}-\ref{fig:2DG12U18})).
The impurity screening is found to $W(i\omega_{m}=0)=16.9$
the effective Hubbard ${\cal U}=17.1$, implying  a reduced bandgap.

\noindent

\begin{figure}[H]
\centering
\includegraphics[height=6cm]{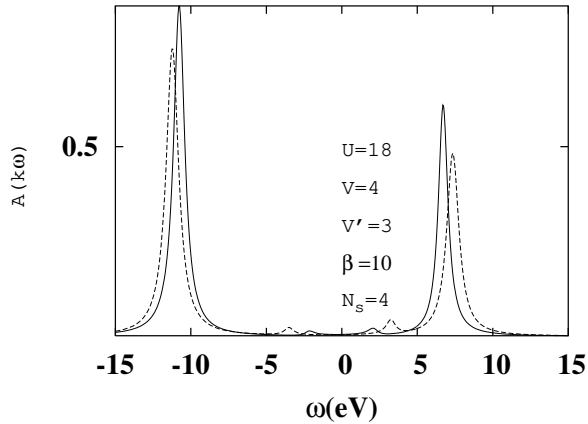}
\caption[]{Spectral function at the $\Gamma$-point.
The chemical potential corresponds to energy zero.
The solid line corresponds to the DMFT+GW case and the dashed line to
the DMFT case.
We used an artificial broadening of 0.5.
%We used 6 Pade coefficients and an artificial broadening of 0.5.
Parameters used: $N_{k}=169$, $N_{g}=512$, $N_{h}=128$ and $N_{P}=N_{S}=64$.
} \label{fig:2DAkwU18}
\end{figure}

The corresponding spectral function is shown
in Fig. (\ref{fig:2DAkwU18}).
We note that in the strong insulator case the Hubbard gap ($\sim {\cal U}$)
is indeed somewhat reduced due
to the inclusion of the GW self-energy.

%\noindent
%\begin{figure}[H]
%\begin{center}
%\centerline{
%\includegraphics[height=6cm]{2DAkwU18VV.ps}}
%\vspace{0.5cm} \caption[]{
%Spectral function at the $\Gamma$-point.
%The chemical potential corresponds to energy zero.
%The solid line corresponds to the $V=4$ and $V'=3$ case and the dashed line to
%$V=V'=0$. We used 6 Pade coefficients and an artificial
%broadening of 0.5.
%Parameters used:
%$N_{k}=169$, $N_{g}=512$, $N_{h}=128$ and $N_{P}=N_{S}=64$. }
%\label{fig:2DAkwU18VV}
%\end{center}
%\hfill
%\end{figure}

\section{Concluding remarks}

%A fully self-consistent DMFT electronic structure calculation in
%combination with the DFT-LDA counter-part is numerically very
%demanding. For a fix impurity charge density
%$n({\bf r})$
%the one-electron Hamiltonian is defined, including an explicit
%Hubbard-like term for the correlated orbitals as well as the
%double counting term, which can be solved using the DMFT
%approximation.
%The impurity Green's function and
%self-energy is then evaluated.
%(these quantities are $(2l+1)^{2}$ matrices).
%With the aid the (input) one-electron Hamiltonian and the (output)
%impurity self-energy  obtained from the (inner) DMFT loop, an
%updated impurity charge density
%$n({\bf r})$ i
%is evaluated. This
%(outer) cycle is then repeated until the impurity charge density
%$n({\bf r})$
%has converged as well. In a simplified scheme, the
%DFT-LDA calculation is only performed once, and the converged
%one-electron Hamiltonian is feeded into the DMFT loop.
In the present study a full self-consistency is performed,
including the {\em non-local} GW self-energy, in the local single
site DMFT approach and the applicability of the method is tested
for a model system. Eventually at self-consistency the {\em full}
self-energy and polarization operator are obtained, from which e.g
the full screened interaction is accessible. Far from the
metal-insulator transition the combination of the GW method and
the single site DMFT is from a numerical point of view fast and
stable, even when a simple linear mixing scheme is utilized.
Changes with respect to DMFT are in some cases substantial, and
are related to the long-rangeness of the GW kernel, specified by
two hopping parameters.

Next we will study the 2D metal-insulator transition as well as
doping away from half-filling.

\section{Acknowledgments}
We greatly acknowledge discussions with Ferdi Aryasetiawan, Silke
Biermann
 and
Roger Bengtsson.

%\begin{appendix}
%\end{appendix}

%} %large

\begin{thebibliography}{99}
\bibitem{note2} The 2D case is described, however to consider the
1D case the modifications are minor. The hopping matrix have
non-diagonal elements $2cosk$ in the chain case.
\bibitem{kramer}
The Kramer-Kronig (KK) relations can be written as
\begin{eqnarray}
{\rm Re} \Sigma(\omega) =
-\frac{1}{\pi}\int d\omega' \frac{{\rm Im} \Sigma(\omega')}
{\omega-\omega'}
\\
{\rm Im} \Sigma(\omega) =
 \frac{1}{\pi}\int d\omega' \frac{{\rm Re} \Sigma(\omega')}
{\omega-\omega'}
\nonumber
\end{eqnarray}
which  are very important in connecting the real and imaginary part of
a given complex function.
With the knowledge of the self-energy on the whole real axis on can perform
an analytical continuation to the imaginary axis
\begin{eqnarray}
\Sigma(z) =
-\frac{1}{\pi}\int d\omega' \frac{{\rm Im} \Sigma(\omega')}{z-\omega'}
\nonumber
\end{eqnarray}
for any $z$ in the upper half-plane. It is readily shown that
${\rm Re} \Sigma(i\omega)=0$
and ${\rm Im} \Sigma(i\omega) = -{\rm Im} \Sigma(-i\omega)$
if ${\rm Im} \Sigma(\omega)$ symmetric
around $\omega=0$. Furthermore it can be shown that the slopes of
${\rm Re} \Sigma(\omega)$ and ${\rm Im} \Sigma(i\omega)$
are equal for small
$\omega$ and ${\rm Im} \Sigma(i\omega)\sim \delta(\omega)$ if
the spectral function has zero amplitude at the chemical potential
$\omega=0$.
\end{thebibliography}
\end{document}